\documentclass[12pt]{article}
\usepackage[a4paper, mag=1000, includefoot,
           top=2cm, bottom=3cm, left=1.5cm, right=2cm, 
            footskip=1cm]{geometry}
\usepackage[cp1251]{inputenc}
\usepackage{indentfirst}
\usepackage{amssymb} 
\usepackage{amsmath}
\usepackage{eucal}
\usepackage{graphicx}
\usepackage{psfrag}
\usepackage{caption}
\usepackage[numbers,square,comma,compress]{natbib}

\usepackage{ifpdf}

\usepackage{hyperref}

\allowdisplaybreaks

\usepackage{cancel}
\DeclareMathOperator{\sign}{sign}
\DeclareMathOperator{\Sg}{Sg}

\DeclareMathOperator{\Cos}{Cos}
\DeclareMathOperator{\Sin}{Sin}

\newcommand{\om}{\omega}

\newcommand{\VSp}[1]{\rule[#1ex]{0ex}{0ex}}
\newcommand{\HSp}[1]{\mspace{#1mu}}

\newcommand{\vph}{\varphi}
\newcommand{\Zt}{\mathrm Z}
\newcommand{\Ch}[3][\dss]{\lefteqn{#1#2}\hphantom{#3}} 

\newcommand\tp{\tilde{\psi}}
\newcommand{\be}{\begin{equation}}
\newcommand{\ee}{\end{equation}}
\newcommand{\beqa}{\begin{eqnarray}}
\newcommand{\eeqa}{\end{eqnarray}}

\newcommand{\pd}{\partial}

\renewcommand\r{\rho}

\renewcommand\a{\alpha}
\renewcommand\b{\beta}

\newtheorem{Def}{Definition}

\newcommand{\dss}{\displaystyle}
\newcommand{\txs}{\textstyle}
\newcommand{\scs}{\scriptstyle}
\newcommand{\sss}{\scriptscriptstyle}

\newcommand{\ra}{\rightarrow}

\newcommand{\ts}{\times}

\newcommand{\ms}{\mathstrut}
\newcommand{\intl}[2]{\int\limits_{#1}^{#2}}
\newcommand{\tintl}[2]{{\txs \intl{#1}{#2} }}

\newcommand{\const}{\mathrm{const}}

\newcommand{\cm}[1]{}
\newcommand{\ed}{\end{document}}

\newcommand{\al}{\alpha}

\newcommand{\gm}{{\gamma{}}}
\newcommand{\Gm}{{\Gamma{}}}

\newcommand{\Dl}{\Delta}
\newcommand{\ep}{\varepsilon}
\newcommand{\deps}{{\dss\varepsilon}}
\newcommand{\zt}{\zeta}

\newcommand{\Up}{\Upsilon}

\newcommand{\Om}{\Omega}

\newcommand{\RR}{\mathbb{R}}

\newcommand{\tg}{\mathrm{tg}}

\begin{document}

\title{{\huge Asymptotic solutions in f(R)-gravity}}

\author{
  Evgeny E. Bukzhalev,\!$^{a}$\footnote{E-mail: bukzhalev@mail.ru} Mikhail M. Ivanov,\!$^{a,b,c}$\footnote{E-mail: mm.ivanov@physics.msu.ru} and Alexey V. Toporensky,\!$^{c}$\footnote{E-mail: atopor@rambler.ru}\vspace{.2cm}
\vspace{.2cm}\\
\normalsize\llap{$^a$}\it Faculty of Physics, Moscow State University,\\
\normalsize \it Vorobjevy Gory, 119991 Moscow, Russia\\
\normalsize\llap{$^b$}\it
Institute for Nuclear Research of the
Russian Academy of Sciences, \\ 
\normalsize \it  60th October Anniversary Prospect, 7a, 117312
Moscow, Russia \\
\normalsize\llap{$^c$}
 \it Sternberg Astronomical Institute, Moscow State University,\\
 \normalsize\it Universitetsky prospect, 13, 119992 Moscow, Russia
}

\date{}
\maketitle
\begin{abstract}

We study cosmological solutions in $R + \beta R^{N}$-gravity for an isotropic Universe filled with ordinary matter with
the equation of state parameter $\gamma$. 
Using the Bogolyubov-Krylov-Mitropol'skii averaging method we 
find asymptotic oscillatory solutions in terms of new functions,
which have been specially introduced by us for this problem and appeared as a natural generalization of the usual sine and cosine.
It is shown that the late-time behaviour of the Universe in the model under 
investigation is determined by the sign of the difference $\gamma-\gamma_{crit}$ where $\gamma_{crit}=2N/(3N-2)$. If $\gamma < 
\gamma_{crit}$, the Universe reaches the regime of small oscillations near values of Hubble parameter and matter density,
corresponding to General Relativity solution. 
Otherwise higher-curvature corrections become important at late times. 
We also study numerically basins of attraction for the oscillatory and phantom solutions, which are present in the theory for $N>2$.
Some important differences between $N=2$ and $N>2$ cases are discussed.
\end{abstract}

\newpage
\tableofcontents

\section{Introduction}

Theories of modified gravity (see \cite{MG} for a review), motivated initially
 from Quantum Field Theory recently became a matter of intense investigation
mainly in order to describe the observed accelerated expansion of our Universe (\cite{Obs1},\cite{Obs2}).
It has been shown that one of the simplest modified gravity theories, $f(R)$ gravity can, in principle, explain this experimental
fact without any need of exotic matter, though it appeared to be rather tricky
and requires a specially designed (often without any 
background physical motivation)
form of the function $f$ (see \cite{Star2007}-\cite{Frolov13} and also
reviews on $f(R)$ gravity \cite{Sot}-\cite{ON2}). 
On the other hand, detailed studies of cosmological dynamics in $f(R)$ theories 
have revealed existence of cosmological regimes, which are absolutely incompatible
with the picture of Universe we live in. For example, any theory with power-law
$f(R)=R+\beta R^N$ with $N>2$ has a solution containing a "Big Rip" singularity (see \cite{BigR,BR1,BR2,BR3,BR4,BR5,BR6,BR7,BR8,BR9,BR10,BR11,BR12,BR13} for "Big Rip" singularities and \cite{COA,Muller,Schmidt,CDCT,CTD,Super,fRGB} for their presence in $R^N$ gravity and some extensions). 

Keeping 
this in mind it is reasonable to go back to the initial motivation and formulate a
problem: if we assume that some quantum considerations lead to $f(R)$ theory in
a high-energy regime, can we be sure that such corrections to Einstein gravity
do not spoil well-established facts about cosmological evolution?

As a general $f(R)$ theory has an additional degree of freedom 
(dubbed as a scalaron in \cite{Star1980}),
it is natural to expect that a solution close to Einstein gravity should 
be oscillations near the General Relativity (GR) solution.
It is known that for $R+\b R^2$ theory
the effects coming from quadratic curvature corrections can be represented as
an effective massive scalar field, so cosmological dynamics can be considered
as a combination of a smooth evolution and harmonic oscillations imposed on it.
Dynamics in $R+\b R^2$ theory depends on the equation of state for matter, filling the Universe,
\[
p=(\gm-1)\r\,,\quad \text{where $p$ is pressure and $\r$ is an energy density.}
\]
If the equation of state parameter 
$\gamma<1$, the smooth part of solution coincides with GR behaviour, otherwise
the influence of quadratic curvature correction becomes important at late times (see \cite{Miritzis,Gur&Star}).

This work is devoted to asymptotic oscillatory solutions in 
general $R+\beta R^N$--gravity, which can be relevant for the description of
reheating phase after inflation (see \cite{Kaneda:2010qv,Martin:2013tda,Huang:2013hsb} for inflation
in $R^N$--gravity).
A general power-law case differs from $N=2$ case in several points.
First, the oscillations become anharmonic and can not be represented in
elementary functions. Second, the Big Rip solution absent in the case of $N=2$ appears for any $N>2$. In the present paper we address both problems. Using analytical methods we will describe oscillations and find a critical value of $\gamma$ as a function of $N$, generalizing the $\gamma=1$ condition known for $N=2$ to arbitrary $N>1$. 
After, we use numerics to find a region in the 
initial conditions space starting from
which a trajectory indeed reaches an oscillatory regime and do not fall into a 
Big Rip
singularity.  

It should be pointed out that in $R+\beta R^2$ gravity the mentioned oscillations of $R$ lead to gravitational creation of particles and prevent an overproduction of scalarons (\cite{Star1980}). 
However, in the case of $R+\beta R^N$ this mechanism can not be applied since it is unclear how to evolve through the point $f''(R)=0$, where the sclaron rest mass diverges \footnote{Remind the reader, that the sclaron mass is $M^2(R)=[3f''(R)]^{-1}$ in the WKB regime $|M^2| \gg R^2,R_{\mu\nu}R^{\mu\nu}$.} (\cite{Appleby}). This issue has another face, 
called "non-standard singularity", taking place when a coefficient
in front of a higher derivative in the equations of motion (which is proportional to $f''(R)$) vanishes. 
We will find this problem in our research and propose
a technique to avoid such a difficulty at least at the level of background dynamics.
Note, that in this paper we do not consider energy exchange between scalaron and ordinary (including Dark) matter.
In realistic models where the inflation is driven by higher--order terms, this interaction is necessary for a reheating when the scalaron decays into Dark Matter and Standard Model particles (\cite{PaninGorbunov},\cite{Vilenkin}).

Apart from the technique using in our study, many 
other analytical methods have been recently applied to $R^N$ gravity
(\cite{CDCT},\cite{Basilias},\cite{Amendola}). Small oscillations near several
GR solutions have been studied in \cite{Odin}, however, 
our approach allows to describe oscillations in the regime where 
corrections 
to GR can not be considered as small and can 
even dominate GR dynamics.

Our work is organized as follows. 
In the section (\ref{EF}) we present the model and 
discuss its asymptotic behaviour in the Einstein frame.
Section (\ref{Analyt}) 
contains the analytic study of 
dynamical equations in the Jordan frame in the limit 
of weak couplings of $R^N$ term in the action. 
In the subsections (\ref{Unpert}-\ref{Trans}) 
we prepare the equations of motion 
for the Bogolyubov-Krylov-Mitropol'skii averaging procedure
and apply it in the subsection (\ref{Aver}).
The reader, who is not interested in technicalities may go directly to the subsection (\ref{Res}) containing main analytic results. 
In the section (\ref{Num}) we discuss what happens in the case of strong couplings, where the analytical scheme breaks down. 
Section (\ref{conclusions}) contains summary of our results.
Finally, in the Appendix (\ref{App1}) we present a systematic treatment of generalised trigonometric functions, used in (\ref{Analyt}).

\section{The model and a picture in the Einstein frame}
\label{EF}
\begin{figure}[!h]
\begin{center}
\includegraphics[width=0.5\textwidth]{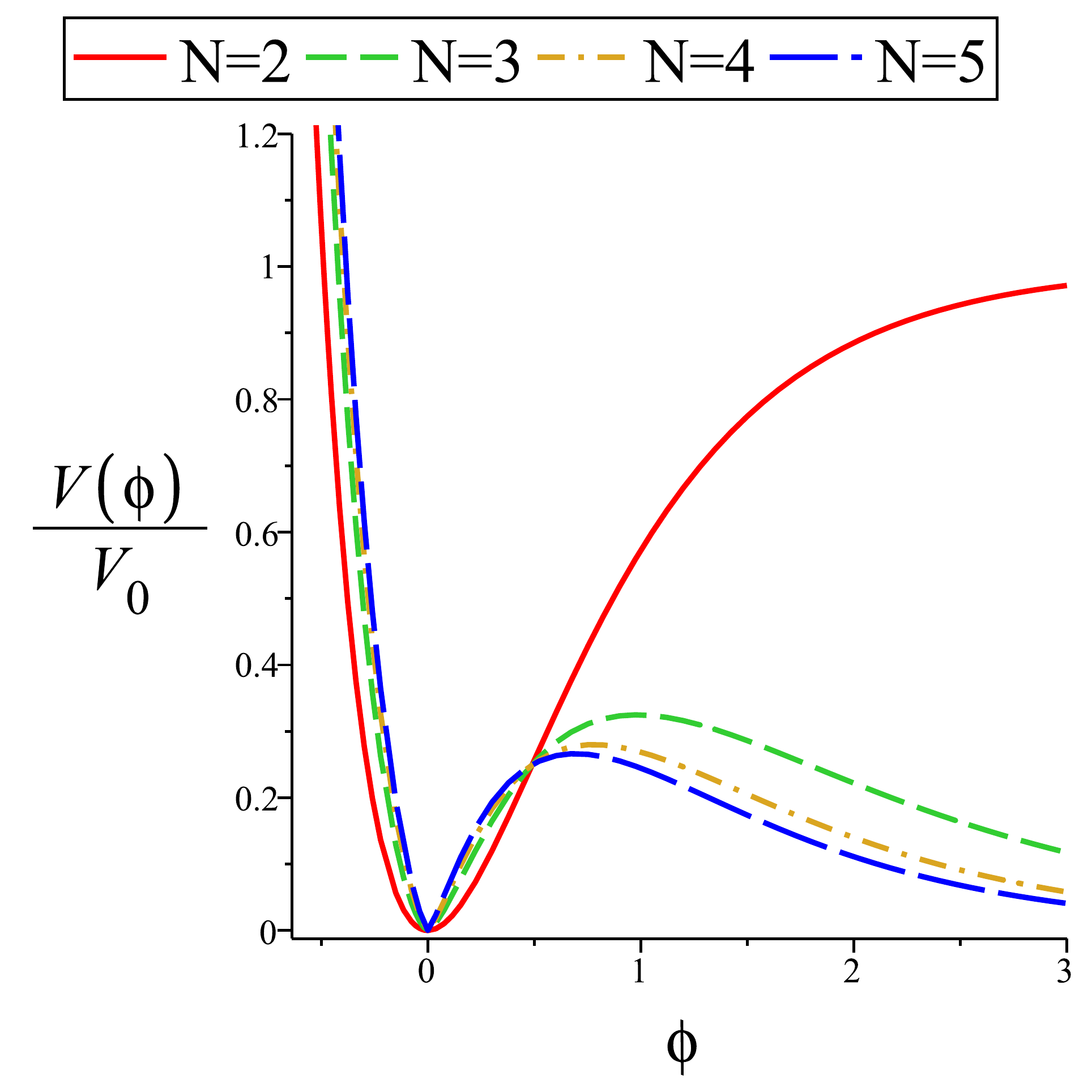}
\caption{Scalaron potential (\ref{Pot}) in the Einstein frame for the function $f(R)=R+\beta R^{N}$. 
\label{Fig:pot}
}
\end{center}
\end{figure}

The action we study 
is given in the Jordan frame by\footnote{
\label{f1}The signature of the metric is assumed 
to be $(-,+,+,+)$, $c=1=\varkappa^2/3=8\pi G/3$.}
\be
\label{actionJF}
S^{JF}=\int d^4x \,\sqrt{-g}   \frac { f(R)}{6}+ S_{matter}^{JF}(g^{\mu\nu},\varphi_i), 
\ee
where $S_{matter}$ is the action for matter fields $\varphi_i$ universally coupled to the metric.
It is well-known that $f(R)$-gravity is classically equivalent to the 
scalar-tensor gravity via the Legendre-Weyl transformation \cite{LWt}.
Let us firstly discuss the picture in the Einstein frame; 
it will help us to understand qualitatively an asymptotic behaviour in the
considered theory.
Performing a standard transition to the Einstein frame (see for instance \cite{fR}) and introducing a scalar degree of freedom (scalaron hereafter) through
\be
\label{JEt}
f'(R)=\exp{[\sqrt{2}\phi]} \,,\quad \text{where}\quad f'\equiv \partial f/\partial R\,,
\ee
one writes down an equivalent action in the form 
\be
S^{EF}= \int d^4x \,\sqrt{-g_E}\,\left[ \frac {R_E}{6}-\frac{1}{2}g^{\mu\nu}\partial_\mu\phi\partial_\nu\phi 
-V(\phi)\right]+S_{matter}^{EF}(g^{\mu\nu}_E,\varphi_i,\phi)\,,
\ee
where the subscript $E$ corresponds to quantities in the Einstein frame.
Notice that the scalaron $\phi$ now
couples to the matter fields.
In the present paper we study the simplest $f(R)$ theory, namely $R+\beta R^N$. This theory has been well-investigated in the Einstein frame in the context of inflationary dynamics (\cite{Kaneda:2010qv,Martin:2013tda,Huang:2013hsb}).
The scalaron potential in the chosen unit system (footnote \ref{f1}) has the following form,
\be
\label{Pot}
V(\phi)= V_0(1-e^{-\sqrt{2}\phi})^{\frac{N}{N-1}}e^{-\frac{N-2}{N-1}\sqrt{2}\phi},\quad\text{where}\quad V_0\equiv \frac{N-1}{6N}\left(\frac{1}{\beta N}\right)^{\frac{1}{N-1}}\,,
\ee
and is plotted on the graph (\ref{Fig:pot}) for a set of parameters $N$.

We clearly see that the case of $N=2$ (Starobinsky model) 
sufficiently differs from other cases of $N$.
The potential for Starobinsky model has smooth non-zero
constant asymptotic
in the limit $\phi\to \infty$, while for other cases of $N$ we have a runaway potential.
As a consequence, the theory with $N>2$ exhibits stable 
singular behaviour, mentioned in the introduction. From the shape of the potential
it is clear that if the scalaron is at the right sight of the potential maxima,
then it would roll down into the area $\phi \to \infty$. Such
behaviour of the scalaron can be translated into
terms of curvature in the Jordan frame 
via (\ref{JEt}), 
\[
\phi=2^{-1/2}\ln{f'}=2^{-1/2}\ln{(1+N\beta R^{N-1})}\to \infty,
\]
implying $R\to \infty$, what indeed represents the known Big Rip singularity of $R^N$-gravity\footnote{The correspondence of singularities between the two frames is a non-trivial problem, which 
we leave for a future work.}. 

In the present paper we are going to investigate attractor solutions 
(i.e. stable regimes occurred at late times)
in the general $R+\beta R^N$ theory. 
From the Fig.\ref{Fig:pot} it is evident that for $N=2$
we have the only oscillating solution near the minimum of the potential
while for $N>2$ two regimes are possible: either scalaron falls into the
minimum of the potential and oscillate there, or rolls down into the unrestricted area located  to the 
right from the maximum of the potential. 
We will describe analytically the oscillatory solution for arbitrary $N$
and generalize the results obtained for the Starobinsky model.
After, we will specify which initial conditions lead to the mentioned 
above two different behaviours.
We choose to work in the Jordan frame for two reasons. First, it is more 
related to the observations (\cite{Nitti:2012ev}) and second for the sake 
of relative simplicity, since the cosmological dynamics in the Einstein frame
for the non-vacuum case 
appeared to be more complicated than the dynamics in the Jordan frame due to 
presence of a scalaron-matter coupling.



\section{The case of weak couplings: analytical study}
\label{Analyt}

In this section we consider the case of $\b \ll 1$, leaving the strong
coupling limit $\b \sim 1$ for the next section. 
The chosen case allows us to study the model analytically and 
find asymptotic oscillatory solution.

Varying the action (\ref{actionJF}) on metric we write down the (00)-equation of motion in $f(R)$-gravity for the flat 
Friedmann-Lemaitre-Robertson-Walker (FLRW) Universe filled with matter with the energy density $\rho$ and
the pressure $p=(\gamma-1)\r$  
:
\begin{align}
\label{eqm}
3f'H^2=(f'R-f)/2-3H\dot{f'}+ 3\rho\,,
\end{align}
supplemented by the Ricci scalar expression and the continuity equation:
\begin{align}
\label{Ricci}
& R=6(2H^2+\dot{H})\,,\\
\label{conteq}
& \dot{\rho}+3H\gamma\rho=0\,.
\end{align}
Using $\dot{f'}=f''\dot{R}$, the eq.\eqref{eqm} may be resolved with respect to the higher derivative
\be 
\label{00f2}
18f''(H,\dot{H})H\ddot{H}=\frac{1}{2}(f'-fR)-3f'H^2-72f''H^2\dot{H}\,+3\r.
\ee
We intend to study the power-law theory $f(R)=R+\beta R^N$.
Looking at eq. \eqref{00f2} one sees that if the scalar curvature $R$ change its sign,
the coefficient $f''$ in front of the higher derivative $\ddot H$
in the case of odd $N$ also change its sign, providing so-called 
"non-standard singularity" (see discussion below). 
To prevent this we insert a modulus in our definition of $f(R)$\footnote{Note, that the "non-standard singularities" are not totally fixed yet, since the coefficient in front of the higher derivative may vanish. We will discuss this issue in Sec. \ref{Num}.}:
\begin{equation*}
f(R)=R+\beta |R|^N\,,\quad f'=1+\beta N |R|^{N-1}(\text{sign} R)\,.
\end{equation*}
Generally speaking, the power index may have any value $N>1$ (including non-integer). 
The (00)-equation then can be rewritten as:
\begin{align}
\label{00before}
H^2=\rho +\frac{\beta(N-1)}{6}|R|^{N-2}\left[R^2-\frac{6N}{N-1}H^2R-6HN\dot{R} \right] \,
\equiv \rho + \beta \rho_{f(R)}\,.
\end{align}
The sign of coupling $\beta$ is fixed to be positive by the stability conditions $f'>0, f''>0$ (\cite{Nunez,Sawicki,Dolgov}), which also exclude the case of $N<1$ from our consideration.
In general the eq. (\ref{00before}) is a non-linear 
second order differential equation, which is impossible to solve explicitly.
However, in the case of small coupling $\b$ in front of the higher 
derivative term,
one can find an asymptotic solution (with respect to the constant $\b$). 
Introducing some useful notations:
\begin{align}
\varepsilon^N\equiv(N-1)6^{N-1}\beta\,,\quad R=\frac{6Q}{\varepsilon}\,, \nonumber \\
\tau=\frac{t-t_0}{\varepsilon} \quad \Rightarrow \quad t=t_0+\varepsilon \tau\,,
\end{align}
we find the following expression for the second term in eq.(\ref{00before}):
\begin{align}
\beta\rho_{f(R)}=-|Q|^{N-2}(NHQ_{\tau}-Q^2+\frac{N}{N-1}\varepsilon H^2 Q)\, \quad \text{where} \quad Q_{\tau}\equiv \frac{dQ}{d\tau}\,.
\end{align}
Finally, collecting eqs. (\ref{Ricci},\ref{conteq},\ref{00before}) one finds the following system of equations:
\begin{subequations}
\label{system1}
\begin{align}
\label{sys1eq1}
H_{\tau}&=Q-2\varepsilon H^2\,,\\
\label{sys1eq2}
NH|Q|^{N-2}Q_{\tau}&=|Q|^N+\rho-H^2-\frac{N}{N-1}\varepsilon H^2Q|Q|^{N-2}\,,\\
\label{sys1eq3}
\rho_{\tau}&=-3\varepsilon\gamma H\rho\,.
\end{align}
\end{subequations}
General solution of this system as well as its integrals (finite expressions, defining a solution implicitly) can not be expressed in quadratures. Our goal is to find an asymptotic (regarding the small parameter $\varepsilon$) approximate solution, i.e. the solution with a precision, growing unlimitedly with $\varepsilon$ tending to zero. 

Let us discuss our strategy. We intend to find an asymptotic solution of the system (\ref{system1}) using the Krylov-Bogolyubov-Mitropol'skii averaging method (\cite{BogKril}). To do that one should change variables and transform initial equations to the so-called standard form, which may be of the
two different variates: either the usual one or the system with rapidly rotating phase \cite{BogMitr}. We will find that the last one is the case for our model;
the resulting system will take a form
\begin{align}
\label{raprot}
 \Bigg\{
   \begin{aligned}
   \textbf{x}_t&={\bf X}(\textbf{x},\psi,\ep)\,,\\
\psi_t&=\ep^{-1}\omega(\textbf{x})+\Psi({\bf x},\psi,\ep)\,,
  \end{aligned}
\end{align}
where $\bf{X}$,$\Psi$ are periodic functions 
of the phase $\psi$ and $\textbf{x}=(x,y)$ is a vector of new variables.
In this system variables $\bf{x}$ are "slow" in the sense that ${x}_{t}\sim 1$,
${y}_{t}\sim 1$ and 
the variable $\psi$ is "rapid" because $\psi_t\sim \ep^{-1}\gg1$.
Thus for one period of oscillations the 
functions $\bf{X}$,$\Psi$ change slowly in comparison with 
the changing rate of $\psi$, which allows us to average $\bf{X}$,$\Psi$
with respect to $\psi$. 
The presence of a large parameter 
in the r.h.s of a dynamical equation can be considered as a 
definition for systems with rapidly rotating phase (see the excellent monograph \cite{BogMitr} for more
details). 
If it is not the case and the system contains only "slow" variables, 
then it has the so-called usual standard form just as the first line of
\eqref{raprot}. 

To find a proper form of a transformation, we  
put $\ep=0$ in all the equations (\ref{system1}) 
and obtain a general unperturbed solution.
At the next step we vary integration constants of the unperturbed solution
and use them as new variables $\bf{x}$ for the problem (\ref{raprot}).
At the final step we solve the system (\ref{raprot}) by the averaging method.

\subsection{General solution of the unperturbed system}
\label{Unpert}

Following the proposed algorithm, we consider the unperturbed regular system:
\begin{subequations}
\label{systemUnpert}
\begin{align}
\label{1st}
\rho_{\tau}&=0\,,\\
H_{\tau}&=Q\,,\\
NH|Q|^{N-2}Q_{\tau}&=|Q|^N+\rho-H^2\,.
\end{align}
\end{subequations}
A solution of the system (\ref{systemUnpert}) can not be found explicitly as before, but one may find its integrals, i.e. some expressions containing $H,Q$ and $\r$, defining solution implicitly. Solving the first equation \eqref{1st} one immediately finds
\begin{subequations}
\label{systemUnpert2}
\begin{align}
\label{eq1}
\r&=\text{const}\equiv C_2\,,\\
\label{eq2}
H_{\tau}&=Q\,,\\
\label{eq3}
NH|Q|^{N-2}Q_{\tau}&=|Q|^N+C_2-H^2\,.
\end{align}
\end{subequations}
 
Let us multiply eq. (\ref{eq3}) on $Q$ and divide by $H^2$ (which always has non-zero value):
\be
\label{tran1}
\frac{N|Q|^{N-2}QQ_{\tau}}{H}-\frac{|Q|^NQ}{H^2}=\frac{C_2Q}{H^2}-Q 
\ee
Note, that the numerator of the first fraction is exactly $d|Q|^N/d\tau$:
\be
\label{Qder}
(|Q|^N)_{\tau}=N|Q|^{N-1}|Q|_{\tau}=N|Q|^{N-2}(Q\, \text{sign}Q)(Q\, \text{sign}Q)_\tau=N|Q|^{N-2}QQ_{\tau}. 
\ee
Next, dividing (\ref{eq2}) on $H^2$,
\be 
\frac{Q}{H^2}=\frac{H_{\tau}}{H^2}=\left(-\frac{1}{H}\right)_{\tau}\,,
\ee 
and using eqs. (\ref{tran1},\ref{Qder}) one obtains:
\be
(|Q|^N)_{\tau}\left(\frac{1}{H}\right)+(|Q|^{N})\left(\frac{1}{H}\right)_{\tau}=-C_2 \left(\frac{1}{H}\right)_{\tau} - H_{\tau} \,,
\ee
leading to the new integral of motion:
\be
\frac{|Q|^{N}}{H}+H+\frac{C_2}{H}=C=2C_1\,.
\ee
Using (\ref{eq2}) and performing some simple algebraic actions, one finds 
the first-order equation for $H$:
\be
\label{1eqH}
|H_\tau|^N+(H-C_1)^2=C_1^2-C_2\,. 
\ee
To obtain a solution of this equation one has to solve an inversion problem for elliptic ($N=3,4$) or hyperelliptic ($N>4$) integrals \cite{Ellipt}, i.e. one has to find corresponding inverse functions. However, in doing so the solution of \eqref{1eqH} (in terms of mentioned inverse functions) can not be expressed neither in elliptical nor in elementary functions (except the cases of $N=2$ and $C_1^2-C_2=0$).
To proceed we have defined and studied properties 
of new functions, describing the solution;
these functions appeared to be a 
natural generalisation of usual trigonometric sine and cosine. 
The reader may find all the details in the Appendix \eqref{App1};
in the main text we just present the final solution of the unperturbed system:
\begin{subequations}
\label{sol0}
\begin{align}
H&=H(C_1,C_2,\psi)\equiv C_1+\sqrt{C_1^2-C_2}\sin_{N}(\psi)\,,\\
Q&=Q(C_1,C_2,\psi)\equiv \Sg_N(\psi)\sqrt[N]{C_1^2-C_2}\sqrt[N]{1-\sin^2_N(\psi)}\,,\\
\r&=\r(C_1,C_2,\psi)\equiv C_2\,,
\end{align}
\end{subequations}
where the function $\sin_{N}(\psi)$ is defined in (Def. \ref{Def Cos Sin}, \ref{defSinN}) and 
\begin{align}
\label{psiphase}
\psi&\equiv (C_1^2-C_2)^{\frac{2-N}{2N}}\tau+\text{const}\,,\\
\sign\sin'_N(\psi)&=\sign Q(\tau)\equiv \Sg_N(\psi)\,.
\end{align}
From \eqref{psiphase} one observes the appearance of a "rapid phase". 
For $N\neq2$ the phase $\psi$ depends on the integration constants which we 
are going to vary according to the Bogolyubov-Krylov procedure.
As a result, $\psi_\tau$ will stop to be a constant, $\psi_\tau=\psi_{\tau}(C_1(\tau),C_2(\tau),C_1(\tau)_\tau,C_2(\tau)_\tau,\tau)\neq\text{const}$;
and after proper transformations presented below it will 
take a form of the second line in the Eqs.(\ref{raprot}).

\subsection{Transformation to the system with rapidly rotating phase}
\label{Trans}


To solve the perturbed system (\ref{system1}) let us vary the constants $C_1, C_2$
with respect to the time and define $C_1=x(\tau),C_2=y(\tau)$\,. The solution of (\ref{sol0}) then reads:
\begin{subequations}
\label{sysUnperFin}
\begin{align}
\label{H of x y psi}
H&=H(C_1,C_2,\psi)\equiv x+\sqrt{x^2-y}\sin_{N}(\psi)\,,\\
Q&=Q(C_1,C_2,\psi)\equiv \Sg_N(\psi)\sqrt[N]{x^2-y}\sqrt[N]{1-\sin^2_N(\psi)}\,,\\
\label{rho of x y psi}
\r&=\r(C_1,C_2,\psi)\equiv y\,.
\end{align}
\end{subequations}
As it has been discussed, $x(\tau),y(\tau),\psi(\tau)$ are new variables for the problem (\ref{raprot}).
Now let us transform the system (\ref{sysUnperFin}) to the dynamical system for $x,y,\psi$, i.e. let us get equations for
$x_{\tau},\psi_{\tau}$ (an equation for $y_{\tau}$ is already obtained, see (\ref{sys1eq3})). 
Performing a simple differentiating of the function $Q(H,x,y)=\Sg_N(\psi) \sqrt[N]{x^2 - y - {(H - x)}^2}$,
\begin{align*}
  Q_\tau &= Q_H\, H_\tau + Q_x\, x_\tau + Q_y\, y_\tau  = \frac2N\, \frac{x-H}{{|Q|}^{N-2}} + \frac2N\, \frac{H\, x_\tau}{{|Q|}^{N-2}\, Q} - \frac1N\, \frac{y_\tau}{{|Q|}^{N-2}\, Q} - \frac4N\, \ep\, \frac{(x-H)\, H^2}{{|Q|}^{N-2}\, Q}\,,
\end{align*}
 where we defined lower subscripts as
\begin{equation*}
z_\tau\equiv\frac{d}{d\tau}z\,,\quad z_{j}\equiv \frac{\pd}{\pd j}z\,,\quad j=(\psi,x,y)\,,
\end{equation*}
and taking into account the eqs. (\ref{sys1eq2},\ref{sys1eq3}) one finds the following expression for $x_\tau$:
\begin{multline*}
  x_\tau
  = {}- 2\, \ep\, (x^2 - y) + \tfrac{3\, N - 4}{2\, (N-1)}\, \ep\, (x^2 - y)\, (1 - \sin^2_N(\psi)) - 2\, \ep\, x\, \sqrt{x^2-y}\, \sin_N(\psi) {}- \tfrac32\, \ep\, \gm\, y\,.
\end{multline*}

Using the eq.(\ref{sys1eq1}) we write down  
\begin{align}
 & H_\tau = H_x\, x_\tau + H_y\, y_\tau + H_\psi\, \psi_\tau\ = \, Q \, = \sqrt[N]{x^2 - y}\, \sin'_N(\psi) = H_\psi\, {(x^2 - y)}^{\frac{2-N}{2N}}\,,\\
  \label{psiH}
 &\text{and express} \quad \psi_\tau = {(x^2 - y)}^{\frac{2-N}{2N}} - \frac{H_x}{H_\psi}\, x_\tau - \frac{H_y}{H_\psi}\, y_\tau - 2\, \ep\, \frac{H^2}{H_\psi}\,.
\end{align}
Looking at (\ref{sysUnperFin},\ref{sinNprime}) we calculate partial derivatives $H_x,H_y,H_\psi$:
\begin{multline}
\label{mastereq1}
  H_x\, x_\tau + H_y\, y_\tau + 2\, \ep\, H^2 = {}- \frac\ep{2(N-1)}\, \Big\{ 
 - (1 - \sin^2_N(\psi))\, \Big[ (3\, N - 4)\, x^2 + (N - 3\, (N-1)\, \gm)\, y \\
 + (3\, N - 4)\, x\, \sqrt{x^2-y}\, \sin_N(\psi) \Big]
  \Big\}\,,
\end{multline}
and plug them into the eq. (\ref{psiH}):
\begin{multline*}
  \psi_\tau = {(x^2 - y)}^{\frac{2-N}{2N}} - \frac\ep{2\, (N-1)}\, \frac{\Sg_N(\psi)}{\sqrt{x^2-y}}\, \Big[ (3\, N - 4)\, x^2 + (N - 3\, (N-1)\, \gm)\, y +{}
\\
  {}+ (3\, N - 4)\, x\, \sqrt{x^2-y}\, \sin_N(\psi) \Big]\, {(1 - \sin^2_N(\psi))}^{\frac{N-1}N}\,.
\end{multline*}
Going back to the standard physical time $t$,
\begin{equation*}
  \tau = \frac{t - t_0}\ep, \quad t = t_0 + \ep\, \tau, \quad dt = \ep\, d\tau\,,
\end{equation*}
  we obtain the resulting dynamical system to solve:
\begin{subequations}
\label{MasterSyst}
\begin{align}
\label{main system 1}
  x_t &=
  \begin{aligned}[t]
    {}- 2\, (x^2 - y) + \tfrac{3\, N - 4}{2\, (N-1)}\, (x^2 - y)\, [1 - \sin^2_N(\psi)] - 2\, x\, \sqrt{x^2-y}\, \sin_N(\psi) - \tfrac32\, \gm\, y \equiv
  \\
    \equiv X_0(x,y,\psi),
  \end{aligned}
\\
\label{main system 2}
  y_t &= {}- 3\, \gm\, x\, y - 3\, \gm\, y\, \sqrt{x^2-y}\, \sin_N(\psi) \equiv Y_0(x,y,\psi),
\\[1ex]
\label{main system 3}
  \psi_t &=
  \begin{aligned}[t]
    & \ep^{-1}\, {(x^2 - y)}^{\frac{2-N}{2N}} - \tfrac1{2\, (N-1)}\, \tfrac{(3\, N - 4)\, x^2 + (N - 3\, (N-1)\, \gm)\, y}{\sqrt{x^2-y}}\, {[1 - \sin^2_N(\psi)]}^{\frac{N-2}N} \sin'_N(\psi) -{}
  \\
    & {}- \tfrac{3\, N - 4}{2\, (N-1)}\, x\, \sqrt{x^2-y}\, \sin_N(\psi)\, {[1 - \sin^2_N(\psi)]}^{\frac{N-2}N} \sin'_N(\psi) \equiv \ep^{-1}\, \om(x,y) + \Psi_0(x,y,\psi)\,,
  \end{aligned}
\end{align}
where 
\begin{gather*}
  x = x(t,\ep), \quad y = y(t,\ep), \quad \psi = \psi(t,\ep).
\end{gather*}
\end{subequations}
The functions~$X_0(x,y,\psi)$, $Y_0(x,y,\psi)$ and~$\Psi_0(x,y,\psi)$ are periodic with respect to~$\psi$ with the same period~$T_0$ as~$\sin_N(\psi)$\eqref{TsinN}.
The essential fact is that the function at $\ep^{-1}$
in the r.h.s. of~\eqref{main system 3},
\begin{equation}\label{om}
  \om(x,y) = {(x^2 - y)}^{\frac{2-N}{2N}},
\end{equation}
does not depend on~$\psi$. As we have already noted, systems of the form \eqref{MasterSyst} are called systems with rapidly rotating phase (compare with
\eqref{raprot}), 
which is $\psi(t,\ep)$ in our case.

\subsection{First order solution of the perturbed system}
\label{Aver}
Now we are going to solve the system (\ref{MasterSyst}) by the averaging method.
According to the standard scheme, one has to consider the following series expansion for independent variables,
\begin{alignat}{4}\label{separation}
  x &= \xi &&+ \ep\, u_1(\xi,\zt,\varphi) &&+ \ep^2\, u_2(\xi,\zt,\vph) &&+ \cdots, \notag
\\
  y &= \zt &&+ \ep\, w_1(\xi,\zt,\vph) &&+ \ep^2\, w_2(\xi,\zt,\vph) &&+ \cdots,
\\
  \psi &= \vph &&+ \ep\, v_1(\xi,\zt,\vph) &&+ \ep^2\, v_2(\xi,\zt,\vph) &&+ \cdots, \notag
\end{alignat}
where $u_i(\xi,\zt,\vph)$, $w_i(\xi,\zt,\vph)$ and~$v_i(\xi,\zt,\vph)$ are $T_0$-periodic functions of~$\vph$ (see \eqref{TsinN}), which should be expressed from the r.h.s. of~\eqref{MasterSyst}, and
\begin{gather*}
  \xi = \xi(t,\ep), \quad \zt = \zt(t,\ep), \quad \vph = \vph(t,\ep)
\end{gather*}
are new unknown functions, satisfying the averaged system of the form
\begin{alignat}{2}\label{averaged system 1}
  \dot\xi &= \makebox[1em][r]{$\Up_0$}(\xi,\zt) + \ep\, \makebox[1em][r]{$\Up_1$}(\xi,\zt) &&+ \cdots,
\\\label{averaged system 2}
  \dot\zt &= \makebox[1em][r]{$\Gm_0$}(\xi,\zt) + \ep\, \makebox[1em][r]{$\Gm_1$}(\xi,\zt) &&+ \cdots,
\\\label{averaged system 3}
  \dot\vph &= \tfrac1\deps\, \om(\xi,\zt) + \Om_0(\xi,\zt) &&+ \ep\, \Om_1(\xi,\zt) + \cdots.
\end{alignat}
The averaged system~\eqref{averaged system 1}--\eqref{averaged system 3} is much simpler than the sys.(\ref{MasterSyst}) because the equations of "slow motion" \eqref{averaged system 1}--\eqref{averaged system 2} do not depend on $\vph$.
Consequently, one may integrate the 
equations for $\xi(t,\ep)$ and $\zt(t,\ep)$ and use 
them in order to find the function $\vph(t,\ep)$ from the "rapid motion" equation \eqref{averaged system 3}.
In this sense "rapid" and "slow" motions are separated.

To find approximate solution of the system \eqref{averaged system 1}--\eqref{averaged system 3} one has to consider a "shortened" system of the following form :
\begin{alignat*}{2}
  \dot\Xi_n &= \makebox[1em][r]{$\Up_0$}(\Xi_n,\Zt_n) + \ep\; \makebox[1em][r]{$\Up_1$}(\Xi_n,\Zt_n) &&+ \cdots + \ep^n\; \makebox[1em][r]{$\Up_n$}(\Xi_n,\Zt_n),
\\
  \dot\Zt_n &= \makebox[1em][r]{$\Gm_0$}(\Xi_n,\Zt_n) + \ep\; \makebox[1em][r]{$\Gm_1$}(\Xi_n,\Zt_n) &&+ \cdots + \ep^n\; \makebox[1em][r]{$\Gm_n$}(\Xi_n,\Zt_n),
\\
  \dot\Phi_{n-1} &= \tfrac1\deps\, \om(\Xi_n,\Zt_n) + \Om_0(\Xi_n,\Zt_n) &&+ \ep\, \Om_1(\Xi_n,\Zt_n) + \cdots + \ep^{n-1}\, \Om_{n-1}(\Xi_n,\Zt_n),
\end{alignat*}
which is called the averaged system of order $n$ (or the system of $n-1$-th approximation). This system allows us to obtain~$\xi$ and~$\zt$
with an error $O(\ep^{n+1})$ and~$\vph$ with an error~$O(\ep^{n})$:
\begin{equation*}\label{xieq}
  \xi(t,\ep) = \Xi_n(t,\ep) + O(\ep^{n+1}), \quad \zt(t,\ep) = \Zt_n(t,\ep) + O(\ep^{n+1}), \quad \vph(t,\ep) = \Phi_{n-1}(t,\ep) + O(\ep^n).
\end{equation*}
Now consider the first-order averaged system 
\begin{align}
  \dot\Xi_1 &= \makebox[1em][r]{$\Up_0$}(\Xi_1,\Zt_1) + \ep\; \makebox[1em][r]{$\Up_1$}(\Xi_1,\Zt_1), \notag
\\\label{1st order}
  \dot\Zt_1 &= \makebox[1em][r]{$\Gm_0$}(\Xi_1,\Zt_1) + \ep\; \makebox[1em][r]{$\Gm_1$}(\Xi_1,\Zt_1),
\\
  \dot\Phi_0 &= \tfrac1\deps\, \om(\Xi_1,\Zt_1) + \Om_0(\Xi_1,\Zt_1). \notag
\end{align}
By definition, $\Up_0(\xi,\zt)$ is the function $X_0(\xi,\zt,\vph)$
from the r.h.s. of \eqref{main system 1},
averaged on $\vph$:
\begin{align}
\label{Av X_0}
&\Up_0(\xi,\zt) = \frac1T\, \intl0T X_0(\xi,\zt,\vph)\, d\vph \equiv \bar X_0(\xi,\zt),\quad
\text{where } \\
  X_0(\xi,\zt,\vph)& = {}- 2\, (\xi^2 - \zt) + \tfrac{3\, N - 4}{2\, (N-1)}\, (\xi^2 - \zt)\, [1 - \sin^2_N(\vph)] - 2\, \xi\, \sqrt{\xi^2-\zt}\, \sin_N(\vph) - \tfrac32\, \gm\, \zt. \notag
\end{align}
Calculating the integral from \eqref{Av X_0} and using the value $T_0$ from
\eqref{TsinN} one has
\begin{equation*}
  \bar X_0(\xi,\zt) = {}- \tfrac{3\, N}{3\, N - 2}\, (\xi^2 - \zt) - \tfrac32\, \gm\, \zt.
\end{equation*}
Functions $\Gm_0(\xi,\zt)$ and~$\Om_0(\xi,\zt)$ are defined and calculated 
in a pretty similar way:
\begin{alignat*}{2}
  \Gm_0(\xi,\zt) ={} && \bar Y_0&(\xi,\zt) = {}- 3\, \gm\, \xi\, \zt,
\\
  \Om_0(\xi,\zt) ={} && \bar \Psi_0&(\xi,\zt) \equiv 0.
\end{alignat*}
General expressions for $\Up_1(\xi,\zt)$ and~$\Gm_1(\xi,\zt)$ (which we omit here)
have very cumbersome and complex structure, however, it is possible to show after some work that they vanish identically
\begin{equation*}
  \Up_1(\xi,\zt) = \Gm_1(\xi,\zt) \equiv 0.
\end{equation*}
Finally, the system \eqref{1st order} reads:
\begin{align}
  \dot \Xi_1 &= {}- \tfrac{3\, N}{3\, N - 2}\, (\Xi_1^2 - \Zt_1) - \tfrac32\, \gm\, \Zt_1, \notag
\\
\label{1st order ours}
  \dot \Zt_1 &= {}- 3\, \gm\, \Xi_1\, \Zt_1,
\\
  \dot \Phi_0&= \tfrac1\deps\, \om(\Xi_1, \Zt_1). \notag
\end{align}
Integrating the eqs.\eqref{1st order ours}, we find
\begin{gather}
\label{quadr}
  \intl{\Zt_1(t_0)}{\Zt_1(t)}\, \frac1\zt\, {\Big( A\, {\zt\VSp{2}}^{\frac{2N}{(3N-2)\gm}} + \zt \Big)}^{-\tfrac12}\, d\zt = {}- 3\, \gm\, (t - t_0), 
\\\label{slow motion}
  \Xi_1(t) = {\Big( A\, {\Zt_1(t)\VSp{2}}^{\frac{2N}{(3N-2)\gm}} + \Zt_1(t) \Big)}^{\tfrac12},
\\\label{fast motion}
  \Phi_0(t,\ep) = B + \ep^{-1}\, \tintl{t_0}t\, \om(\Xi_1(t), \Zt_1(t))\, dt,
\end{gather}
where~$A$ and~$B$ are the integration constants.
Taking into account~\eqref{om} and~\eqref{slow motion} we get the following expression for~$\om(\Xi_1(t),\Zt_1(t))$:
\begin{equation}\label{om of zt}
  \om(\Xi_1(t),\Zt_1(t)) = {A\VSp{2.2}}^{\frac{2-N}{2N}}\, {\Zt_1(t)\VSp{2.2}}^{\frac{2-N}{(3N-2)\gm}}.
\end{equation}
Next, we plug~\eqref{om of zt} into \eqref{fast motion} and obtain 
\begin{equation}\label{fast motion final}
  \Phi_0(t,\ep) = B + \ep^{-1}\, {A\VSp{2.2}}^{\frac{2-N}{2N}}\, \tintl{t_0}t\, {\Zt_1(t)\VSp{2.2}}^{\frac{2-N}{(3N-2)\gm}}\, dt.
\end{equation}
Eqs. (\ref{slow motion},\ref{fast motion final}) represent 
the first-order solution of 
~\eqref{averaged system 1}--\eqref{averaged system 3} (the solution in zeroth approximation, see the above discussion). Now one may substitute the functions $\Xi_1(t)$, $\Zt_1(t)$ and~$\Phi_0(t,\ep)$ 
into the formulae \eqref{H of x y psi},~\eqref{rho of x y psi}
instead of $x,y,\psi$.
Together with the relations~\eqref{separation} and~\eqref{xieq} this provides us with the zeroth-order approximation for $H$ and the first-order approximation for $\r$:
\begin{equation}\label{approx H rho}
  H(t,\ep) = H_0(t,\ep) + O(\ep), \quad \rho(t,\ep) =  \rho_1(t,\ep) + O(\ep^2),
\end{equation}
where
\begin{align}
     H_0(t,\ep) &= \Xi_1(t) + \sqrt{{\Xi_1(t)}^2 - \Zt_1(t)}\, \sin_N[\Phi_0(t,\ep)], \notag
\\\label{rho1}
  \rho_1(t,\ep) &= \Zt_1(t).
\end{align}
Taking into account~\eqref{slow motion}, the expression for~$H_0$ becomes simpler:
\begin{equation}\label{H0}
  H_0(t,\ep) = \Xi_1(t) + \sqrt A\, {(\Zt_1(t)\VSp{2})}^{\frac N{(3N-2)\gm}} \sin_N[\Phi_0(t,\ep)].
\end{equation}

Finally from (\ref{approx H rho},\ref{rho1},\ref{H0}) the Hubble parameter and the matter energy density as functions 
of time are given by
\begin{align}
\label{finr}
 \rho(t) &= \Zt_1(t)+ O(\ep^2)\,,\\
 \label{finH}
 H(t) &= \Xi_1(t) + \sqrt A\, {(\Zt_1(t)\VSp{2})}^{\frac N{(3N-2)\gm}} \sin_N[\Phi_0(t,\ep)] + O(\ep)\,,
\end{align}
where $\Xi_1(t), \Phi_0(t,\ep)$ should be expressed through (\ref{slow motion}) and (\ref{fast motion}) in terms of the function $\Zt_1(t)$,
which can be found by calculating the quadrature (\ref{quadr}). 
However this quadrature
may be expressed in terms of elementary functions only in the case of 
$\gm=\gm_{crit}$ (see (\ref{gmcrit}) below),
\begin{equation}
\label{quadExact}
  \Zt_1(t) = \frac{C}{{\Big( \tfrac{3\, N}{3\, N - 2}\, \sqrt{C\,(A+1)}\, (t - t_0) + 1 \Big)}^{\!2}},
\end{equation}
where~$C = Z_1(t_0) = \const$.

From the physical point of view it is more interesting to obtain the explicit 
relation between $H$ and $\r$ i.e. the modified Friedmann equation.
According to \eqref{approx H rho} and~\eqref{rho1} :
\begin{equation}\label{approx Zt_1}
  \Zt_1 = \rho + O(\ep^2).
\end{equation}
Using the eqs.\eqref{slow motion}, \eqref{fast motion final},~\eqref{approx Zt_1}, we get
\begin{align}\label{approx Xi_1}
  \Xi_1 &= {\Big( A\, {\rho\VSp{2}}^{\HSp{-0}\frac{2N}{(3N-2)\gm}} + \rho \Big)}^{\tfrac12} + O(\ep^2),
\\\label{approx Phi_0}
  \Phi_0 &= B + \ep^{-1}\, {A\VSp{2.2}}^{\frac{2-N}{2N}}\, \tintl{t_0}t\, {\rho(t)\VSp{2.2}}^{\frac{2-N}{(3N-2)\gm}}\, dt + O(\ep),
\end{align}
which in combination with~\eqref{approx H rho} and~\eqref{H0}--\eqref{approx Zt_1} 
show that 
the Hubble parameter $H$ and the matter energy density 
$\rho$ satisfy the following 
relation:

\noindent $\smash{\rule[-10ex]{.07ex}{10ex}}$ \hrule \hfill $\smash{\rule[-10ex]{.07ex}{10ex}}$
\begin{equation}
\label{Hanalyt}
  H = {\Big( A\, {\rho\VSp{2}}^{\frac{2N}{(3N-2)\gm}} + \rho \Big)}^{\tfrac12} + \sqrt A\: {\rho\VSp{2}}^{\frac N{(3N-2)\gm}}\, \sin_N \Big\{ B + \ep^{-1}\, {A\VSp{2}}^{\frac{2-N}{2N}}\, \tintl{t_0}t\, {[\rho(t)]\VSp{2}}^{\frac{2-N}{(3N-2)\gm}}\, dt \Big\} + O(\ep),
\end{equation}
$\smash{\rule{.07ex}{10ex}}$ \hrule \hfill $\smash{\rule{.07ex}{10ex}}$

where $A,B$ are the integration constants. Note that in the limit $t \to \infty$ $\r \to 0$ as $\r\propto a(t)^{-3\gm}$ due to
the continuity equation (\ref{conteq}). 

The form of the eq.(\ref{Hanalyt}) should not confuse the reader. Indeed, we 
started from the Friedman equation in the form (\ref{00before}) 
containing in the r.h.s two independent energy densities: matter and higher-curvature gravity corrections (HCGC), but 
the eq.(\ref{Hanalyt}) contains the only density of the ordinary matter.
Here the situation can be well understood on the familiar example of 
standard cosmology (based on General Relativity) 
with two types of matter (with the equation-of-state parameters $\gm_1$ and $\gm_2$),
entering in r.h.s of the Friedmann equation
\be
\label{Friedmanusual}
H^2=\r_{1}+\r_{2}\,. 
\ee
According to the continuity equation (\ref{conteq}) these two components 
drop with the scale factor as
\be
\r_{1}=\frac{\r_{1,0}}{a^{3\gm_1}}\,,\quad \r_{2}= \frac{\r_{2,0}}{a^{3\gm_2}}
=\frac{\r_{2,0}}{\r_{1,0}^{\gm_2/\gm_1}}\r_{1}^{\gm_2/\gm_1}\,,
\ee
where in the last equality we have used the scale factor expressed through
the first equation. Thus the Friedmann equation (\ref{Friedmanusual}) can 
be rewritten as 
\be
H^2=\r_{1}+\text{const}\cdot \r_{1}^{\gm_2/\gm_1}\,. 
\ee
Obtaining the eq.(\ref{Hanalyt}) we have used the 
same trick, but in a more complicated manner.
Here the contribution of $\beta \r_{f(R)}=H^2-\r$ is expressed in terms 
of the usual matter density $\r$. Looking at the parentheses in the r.h.s. of (\ref{mastereq1}) it is clear that roughly (neglecting the oscillations, i.e. the second term) 
the effect of HCGC appears in 
a form of additional effective matter with the equation-of-state parameter 
\be
\label{gmcrit}
\gm_{crit} \equiv  \frac{2\, N}{3\, N -2}\,.
\ee

\subsection{Analysis of the oscillating solution}
\label{Res}

Dynamics of the Universe depends on the equation of state parameter $\gm$ and
the value of $\gm_{crit}$ defined in (\ref{gmcrit}),
which give us three following cases:

1) If ~\fbox{$\gm = \gm_{crit} \equiv \frac{2\, N}{3\, N -2}$}\hspace{.1em}, the eq. (\ref{Hanalyt}) yields
\begin{equation*}
  H = \rho^{1/2}\, \Big( \sqrt{A + 1} + \sqrt{A}\, \sin_N \Big\{B + \ep^{-1}\, {A}^{\frac{2-N}{2N}}\, \tintl{t_0}t\, {[\rho(t,\ep)]\VSp{2}}^{\frac{2-N}{2N}}\, dt \Big\} \Big) + O(\ep),
\end{equation*}
In this case it is possible to obtain exact 
analytical expressions 
for the Hubble rate and matter density. Plugging \eqref{quadExact} into 
\eqref{slow motion}, \eqref{fast motion} together with the use of 
\eqref{finr},\eqref{finH} gives us
\begin{equation}
  \rho(t) = \frac{C}{{\Big( \tfrac{3\, N}{3\, N - 2}\, \sqrt{C\,(A+1)}\, (t - t_0) + 1 \Big)}^{\!2}} + O(\ep^2),
\end{equation}
\begin{multline}
\label{Hcrit}
  H(t) = \frac{\sqrt C}{\tfrac{3\, N}{3\, N - 2}\, \sqrt{C\,(A+1)}\, (t - t_0) + 1}\, \Bigg( \sqrt{A+1} + \sqrt A\, \ts
\\%
  {}\ts \sin_N \Bigg\{ B + \ep^{-1}\, \tfrac{3\, N - 2}{6\, (N - 1)}\, \frac{{\big( A\, C \big)}^{\frac{2-N}{2N}}}{\sqrt{C\,(A+1)}}\, \Big[ {\Big( \tfrac{3\, N}{3\, N - 2}\, \sqrt{C\,(A+1)}\, (t - t_0) + 1 \Big)}^{\!\frac{2(N-1)}N} - 1 \Big] \Bigg\} \Bigg) + O(\ep).
\end{multline}
We see that Hubble rate exhibits monotonic decreasing with imposed on it
anharmonic oscillations. The ratio $H^2/\r$ oscillates with the 
constant amplitude which does not equal to unity.

2) In the case of ~\fbox{$\gm < \frac{2\, N}{3\, N -2}$}\hspace{.1em} 
the HCGC energy density decays faster than that of the usual matter,
implying
\begin{equation*}
  \frac{H^2}\rho \ra 1 + O(\ep), \quad t \ra \infty
\end{equation*}

To obtain the result more formally we notice that eqs.\eqref{1st order ours},\eqref{slow motion} give
\begin{equation}
\label{z1eq}
  \dot \Zt_1 = {}- 3\, \gm\, \Zt_1\, {\Big( A\, {\Zt_1\VSp{2}}^{\frac{2N}{(3N-2)\gm}} + \Zt_1 \Big)}^{\tfrac12}.
\end{equation}
Since $Z(t)\to 0$ for $t\to \infty$ ($Z$ is nothing but the matter density $\r$),
then neglecting the first term in the parentheses of \eqref{z1eq} we can integrate a resulting
equation and find approximate expressions for~$\Xi_1$and~$\Phi_0$ (cf. \eqref{slow motion},\eqref{fast motion}). 
This gives us an Einstein solution with damped oscillations, decaying at late times and leading to a pure GR behaviour
\be 
  \rho(t) \approx \frac1{{( \frac32\, \gm\, t )}^2}, \quad H(t) \approx \phantom{(}\frac1{\frac32\, \gm\, t}.
\ee

3) In the last case,~\fbox{$\gm > \frac{2\, N}{3\, N -2}$}, one finds
\begin{equation*}
  H = \sqrt{A}\, {\rho\VSp{2}}^{\frac N{(3N-2)\gm}}\, \Bigg( 1 + \sin_N \Big\{ B + \ep^{-1}\, {A}^{\frac{2-N}{2N}}\, \tintl{t_0}t\, {[\rho(t,\ep)]\VSp{2}}^{\frac{2-N}{(3N-2)\gm}}\, dt \Big\} \Bigg) +  O(\ep),
\end{equation*}
which hints on increasing oscillations of the fraction $H^2/\r$ at late times. 
In its minima $\sin_N(x_{min})=-1$, what gives the estimate:
\begin{multline}
  H \approx {\Big( A {\rho\VSp{2}}^{\frac{2N}{(3N-2)\gm}} + \rho \Big)}^{\tfrac12} - \sqrt{A}\, {\rho\VSp{2}}^{\frac N{(3N-2)\gm}} =
\\
  = \sqrt{A \VSp{-.15}}\, {\rho\VSp{2}}^{\frac N{(3N-2)\gm}} \Big[ {\Big( 1 + \frac1{A}\, {\rho\VSp{2}}^{1-\frac{2N}{(3N-2)\gm}} \Big)}^{\tfrac12} - 1 \Big] \approx \frac12\, \frac1{ \sqrt{A} }\, {\rho\VSp{2}}^{1-\frac N{(3N-2)\gm}}\,.
\end{multline} 
Thus, at points of local minima $H^2\sim {\rho\VSp{2}}^{2 - \frac{2N}{(3N-2)\gm}}\ll \r$ and at points of local maxima $H^2\sim {\rho\VSp{2}}^{\frac{2N}{(3N-2)\gm}}\gg\r$. 
Dynamics is determined by non-linear curvature corrections, becoming dominant at late times. 
Asymptotic expressions for $H$ and $\r$ in the limit $t \to \infty$ can be found analogously to that of the previous case (one only has to neglect the second term in the parenthesis of \eqref{z1eq}),
\begin{equation}
\label{dom_osc}
  \rho(t) \approx {\Big( \tfrac{3N}{3N-2}\, \sqrt A\, t \Big)}^{\!\!-\frac{(3N-2)\gm}N}, \quad
  H(t) \approx {\Big( \tfrac{3\, N}{3\, N - 2}\, t \Big)}^{\!\!-1} \Big\{ 1 + \sin_N \big[ \Phi(t) \big] \Big\},
\end{equation}
where
$$
  \Phi(t) \approx \ep^{-1}\, \tfrac{3\, N - 2}{6\, (N - 1)}\, {\Big( \tfrac{3N}{3N-2}\, t \Big)}^{\!\!\frac{2(N-1)}N}.
$$
One notices that ordinary matter decays with time faster than in the GR cosmology.

Comparing the Friedmann equation \eqref{00before} 
\[
H^2=\r+\beta \r_{f(R)}\,,
\]
with \eqref{Hanalyt} we see that value $\gm_{crit}= \frac{2\, N}{3\, N -2}$ denotes an effective equation of state parameter
for $\b \r_{f(R)}$ and the picture qualitatively coincides with evolution of
the Universe, filled with two types of fluid with the equation--of--state parameters
$\gm$ and $\gm_{crit}$.
For $\gm<\gm_{crit}$ higher curvature terms are sub-dominant at late times and the evolution is driven
by ordinary matter;
for $\gm=\gm_{crit}$ both contributions are equally important
and in the case of $\gm>\gm_{crit}$ higher curvature term dominates the dynamics.
The limit $N\to \infty$ leads to $\gm_{crit}=2/3$, corresponding to
Milne cosmology $a\propto t$ and the case of $N=2$ gives the known result
$\gm_{crit}=1$. 
Therefore these oscillations can not be the cause of an accelerating Universe.
Also it is clear that for $N>2$ the Universe filled with dust ($\gm_{dust}=1$) and radiation ($\gm_{rad}=4/3$) will fall into the oscillatory regime because $\gm_{crit}<\gm_{rad},\gm_{dust}$.

\section{Numerics: the case of strong couplings and basins of attraction for asymptotic solutions}
\label{Num}

\begin{figure}[!h]
\begin{center}
\includegraphics[width=0.45\textwidth]{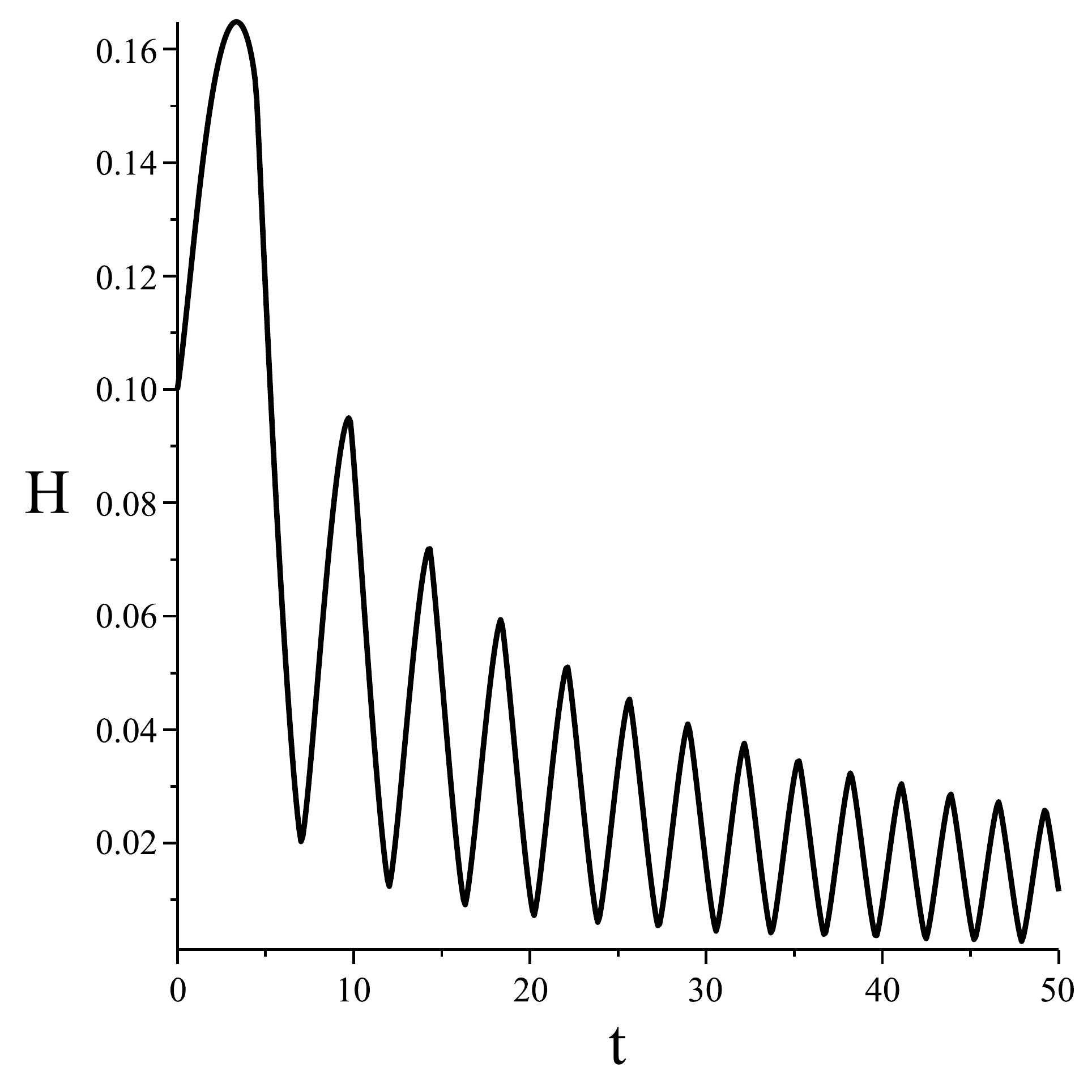}
\includegraphics[width=0.45\textwidth]{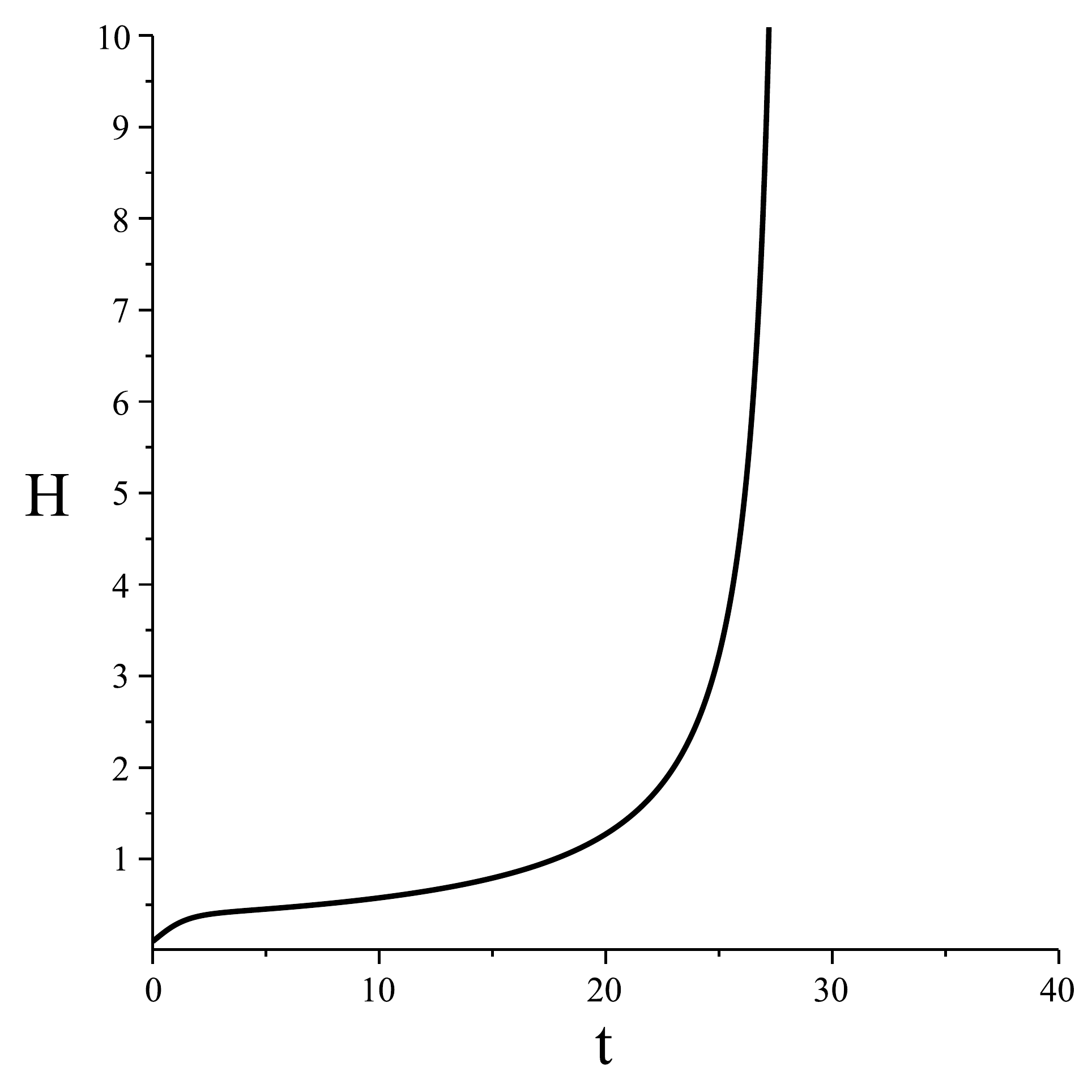}
\caption{Stable asymptotics for $\beta=1, N=4$. The oscillations (left panel) take place for the initial values $(H_0,\dot{H}_0)=(0.1,0.02)$ 
while the runaway singularity occurs
for $(H_0,\dot{H}_0)=(0.1,0.2)$. 
The ordinary matter is dust with $\gamma=1$ and $\r_{0}=0.03$.
The measure units are given in the system, where $8\pi G/3=1$, see
footnote \ref{f1}.
\label{Fig:0}
}
\end{center}
\end{figure}

\begin{figure}[!h]
\begin{center}
\includegraphics[width=0.4\textwidth]{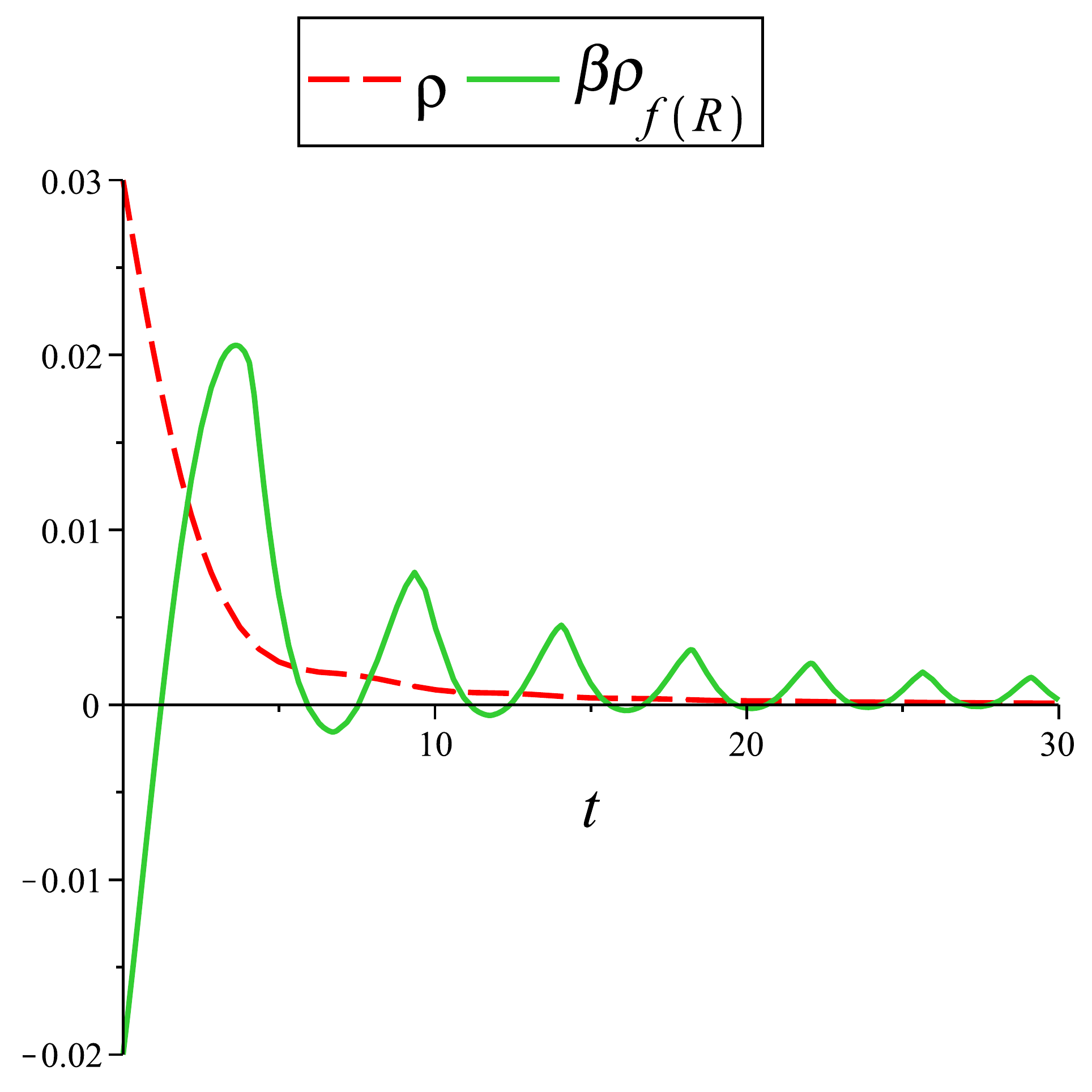}
\includegraphics[width=0.4\textwidth]{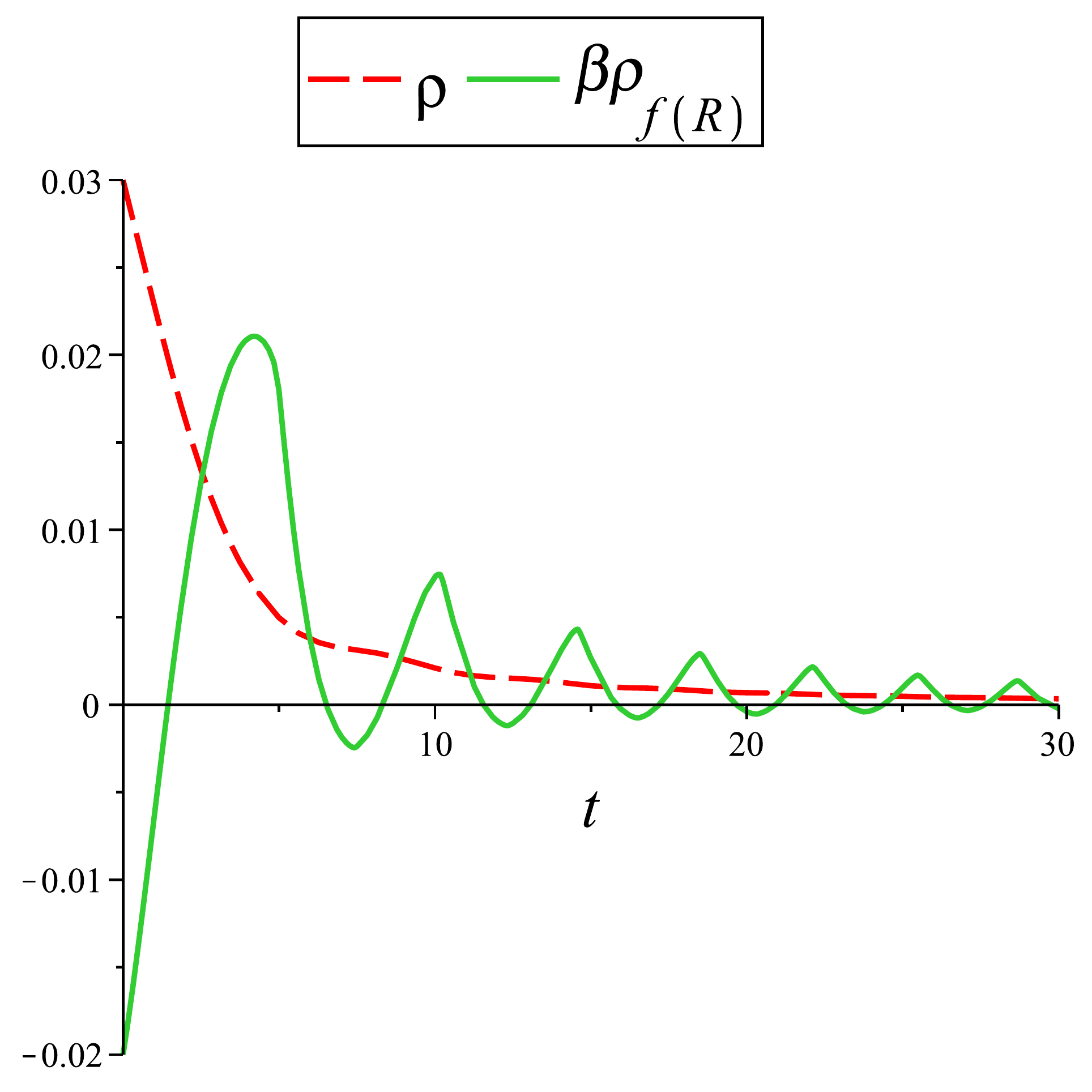}
\includegraphics[width=0.4\textwidth]{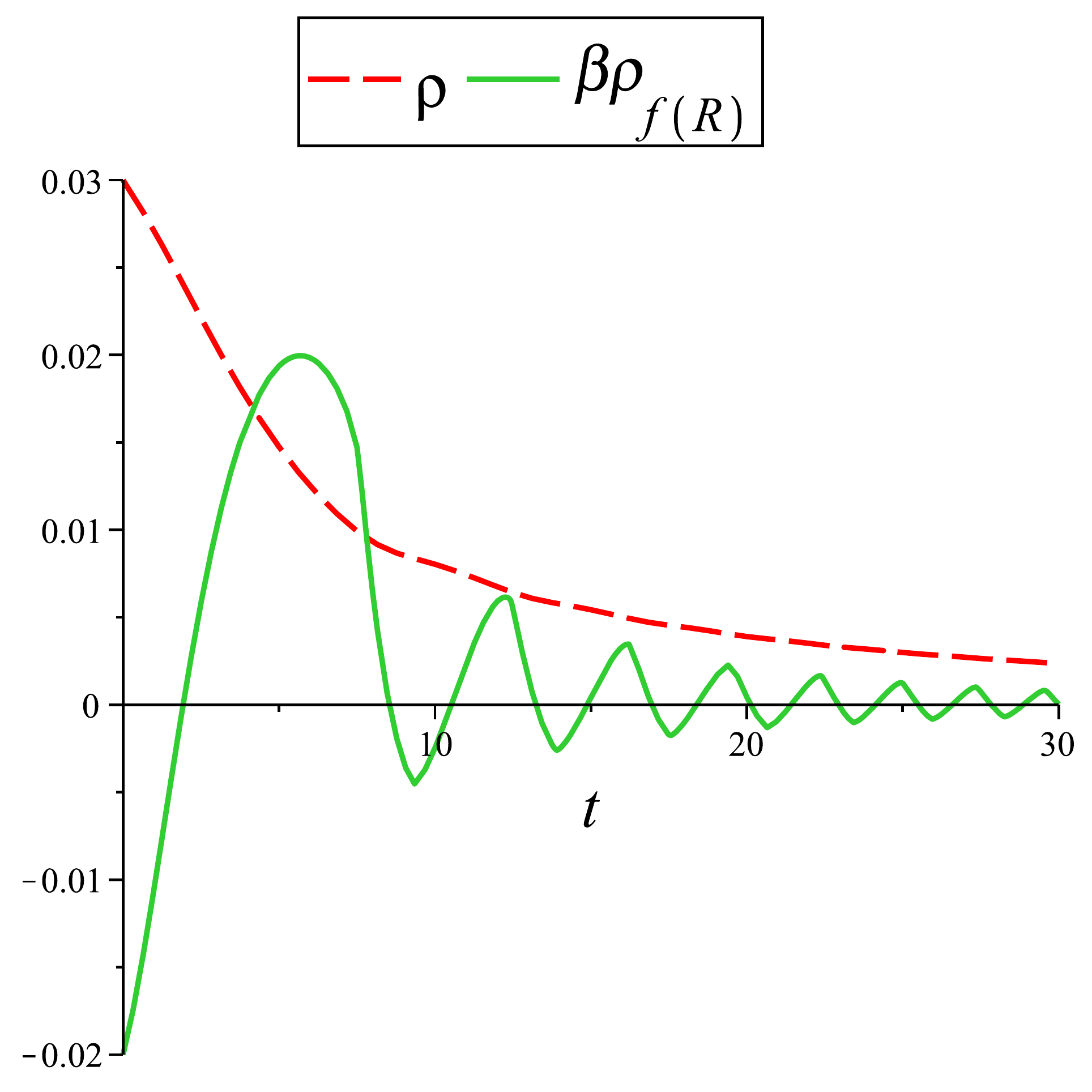}
\caption{Matter energy density (dashed red lines) and higher-curvature corrections
effective energy density
(sold green lines) for an oscillatory solution depending on $\gamma$:
$\gamma>\gamma_{crit}$ (upper left panel), $\gamma=\gamma_{crit}$ (upper right panel),
$\gamma<\gamma_{crit}$ (bottom panel). Model parameters are $\beta=1, N=4\,(\gm_{crit}=0.8)$, initial conditions $(H_0,\dot{H}_0,\r_{0})=(0.1,0.02,0.03)$; $\gm=1.2, 0.8$ and $0.3$, respectively.
\label{Fig:5}
}
\end{center}
\end{figure} 
\begin{figure}[!h]
\begin{center}
\includegraphics[width=0.45\textwidth]{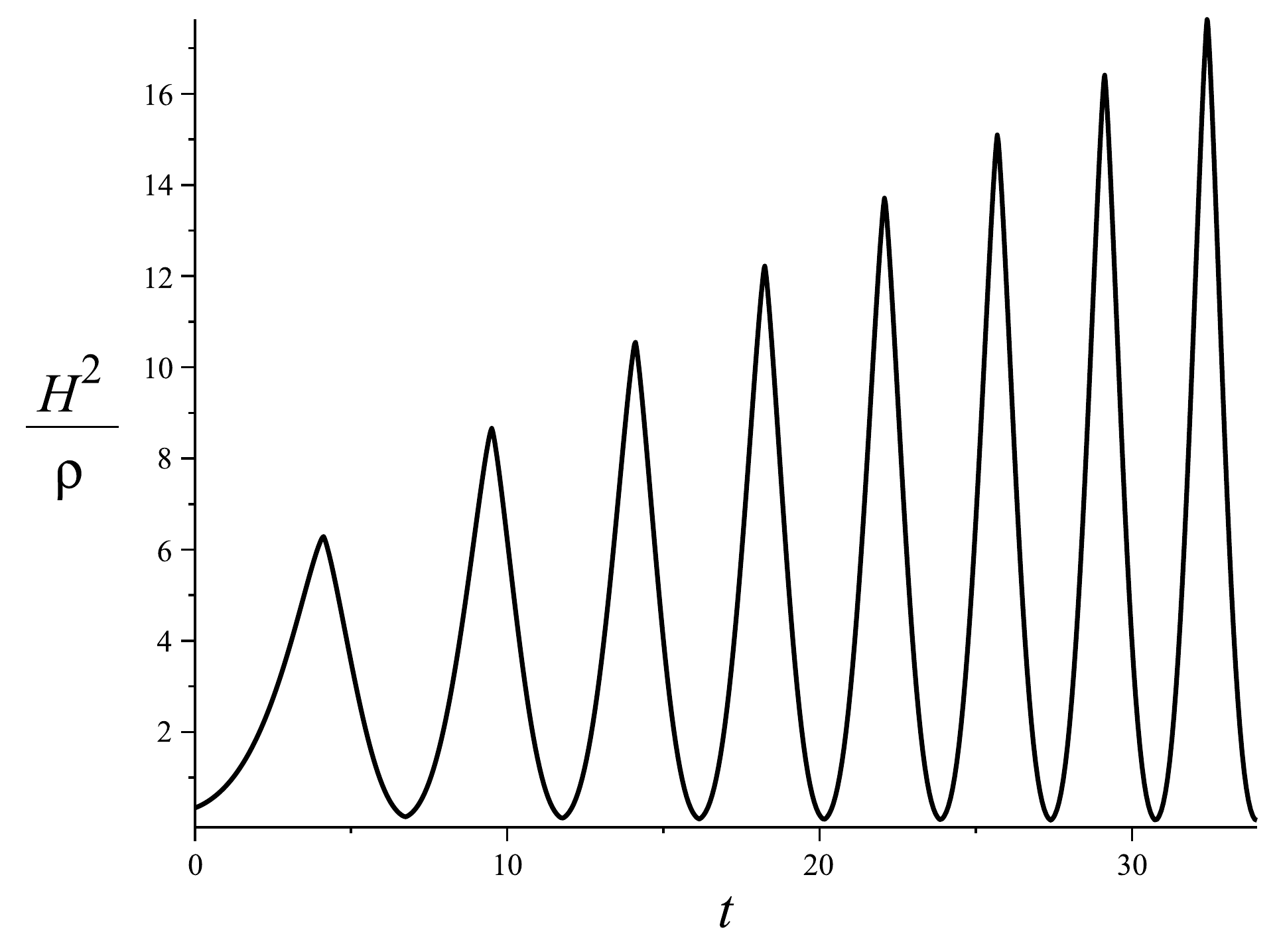}
\includegraphics[width=0.45\textwidth]{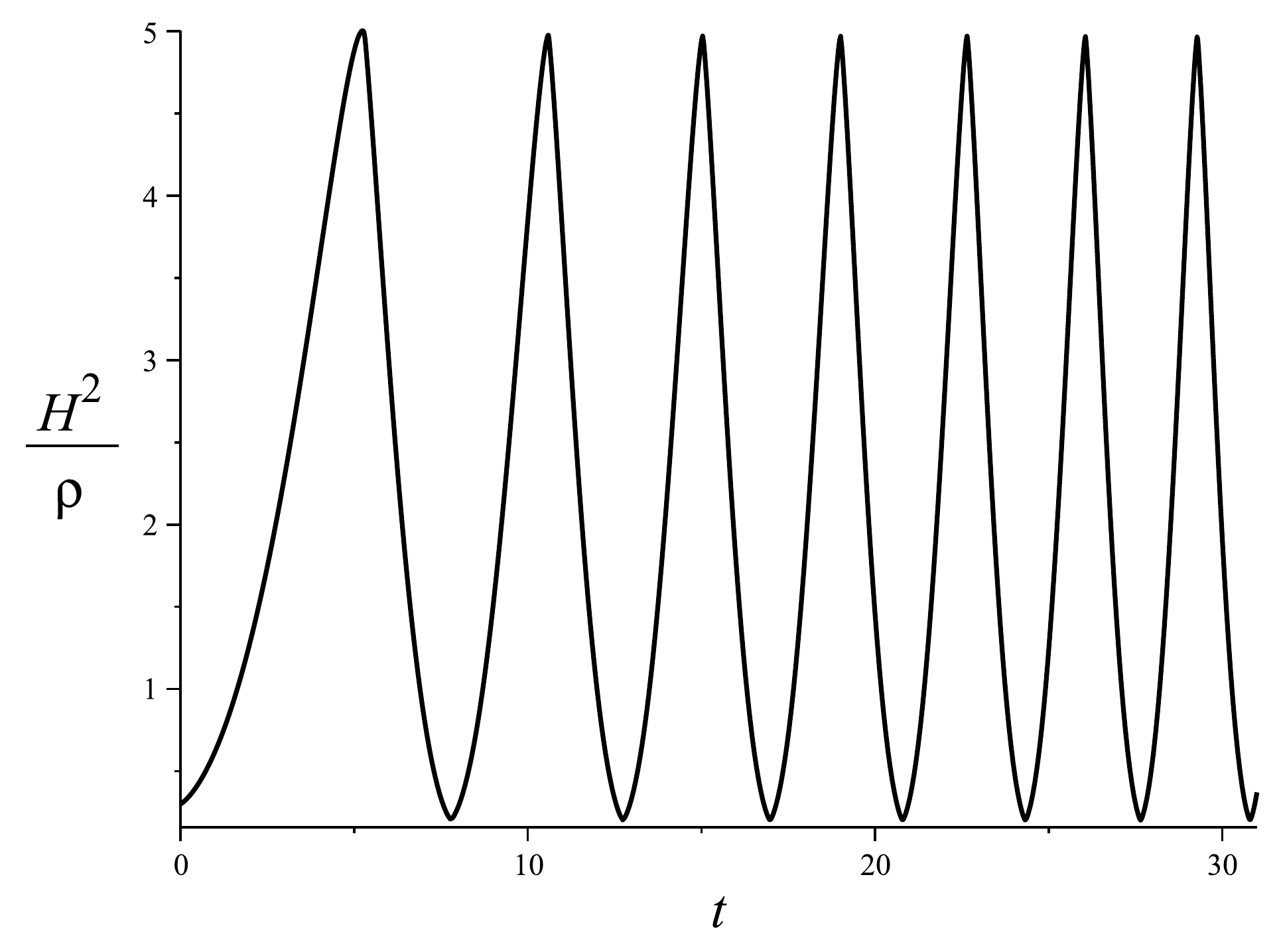}
\includegraphics[width=0.45\textwidth]{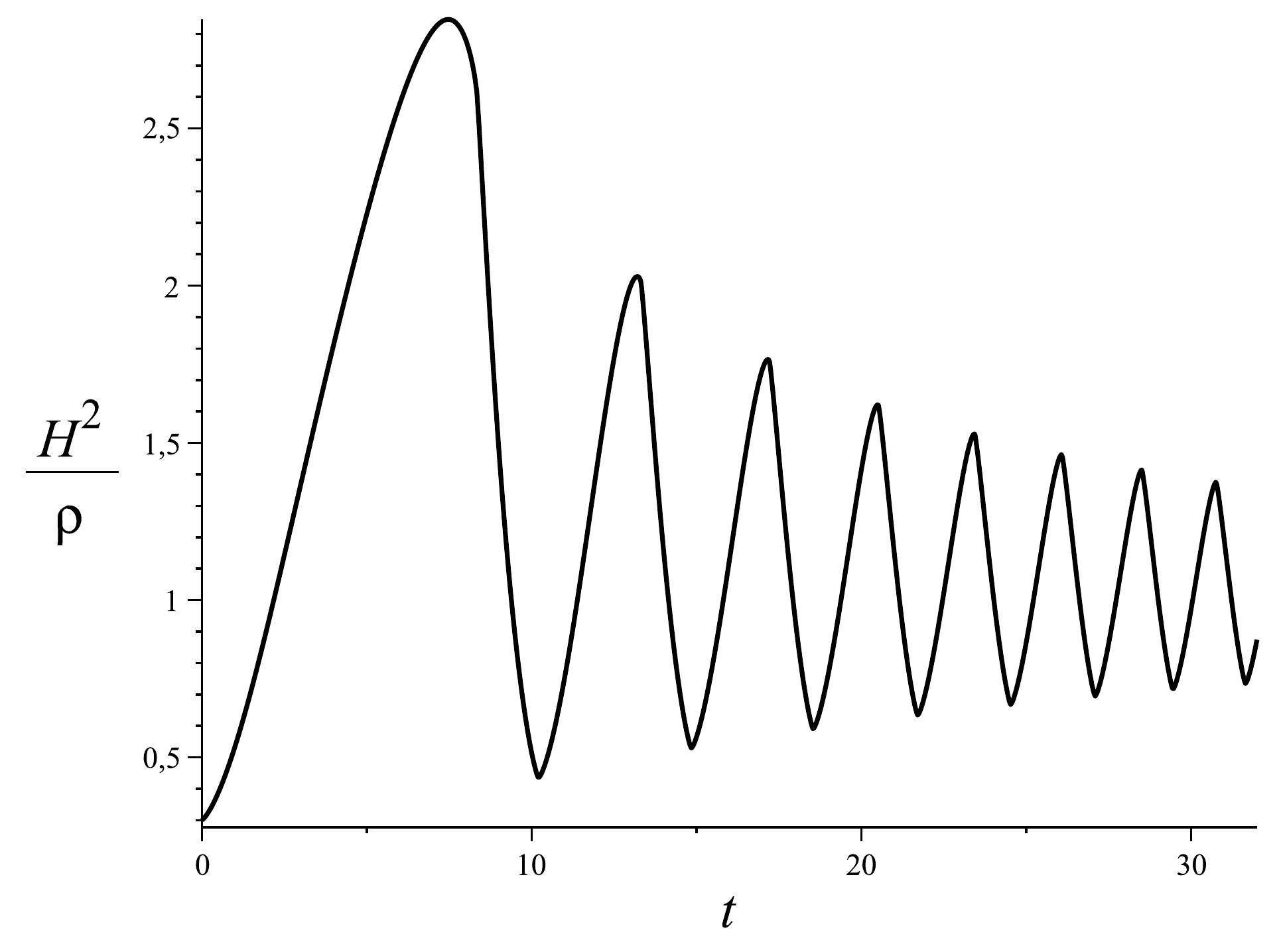}
\caption{Fraction $H^2/\r$ for oscillating solution depending on $\gamma$:
$\gamma>\gamma_{crit}$ (upper left panel), $\gamma=\gamma_{crit}$ (upper right panel),
$\gamma<\gamma_{crit}$ (bottom panel). Model parameters are the same with Fig.(\ref{Fig:5})
\label{Fig:3}
}
\end{center}
\end{figure}

We remind the reader that for a pure $R+\beta R^N (N>2)$ theory the coefficient in
front of the higher derivative ($f''(R)\sim|R|^{N-2}$, see (\ref{00f2})) may vanish for $R=0$.
If this trouble occurres, then $\ddot{H}$ (or, equivalently, $\dot{R}$) diverges in order to satisfy the Eq. (\ref{00f2}), 
leaving $H$ and $\dot H$ finite.
This picture is similar to the so--called
"non-standard singularity" actively studied recently in various generalizations of GR
\cite{Barrow1, Barrow2, Tsagas, brane, 
Mariusz, brane2, Laszlo, Nojiri:2005sx, Nojiri:2004ip, Bamba:2008ut}. 
However, in most of these scenarios the Hubble derivative $\dot H$ diverges, 
keeping $H$ finite, while
in our case the divergence is "shifted" to the second 
derivative of the Hubble rate.
Since $R$, $R_{\mu \nu}R^{\mu \nu}$ and
$R_{\mu \nu \lambda \rho}R^{\mu \nu \lambda \rho}$ 
are expressed in terms of $H$ and $\dot H$ only (see, for example, \cite{Carroll:2004de}), then
this particular situation 
does represent a very weak singularity with 
finite curvature invariants. 

In the analytical part we have neglected this issue\footnote{
This singularity actually exists in the analytical solution,
because $\sin_N''(x)\sim 1/[\cos_N(x)]^{N-2}$ (see \eqref{z''}).
According to \eqref{Hcrit}\eqref{dom_osc} this implies that $\ddot H$ diverges in a discrete set of points.}, 
but it is the crucial problem for a numerical procedure.
However it appeared possible to remove completely this difficulty 
by a small change of the theory 
shifting minimal possible $f''(R)$ by some tiny constant into a zone of positive values.
We have 
added the regularization term $\alpha R^2$ with very 
small coupling $\alpha\ll\beta$ which 
prevents the mentioned issue by making $f''=2\a\neq0$ for $R=0$
and does not modify the dynamics\footnote{
We decreased the value of $\a$ in our numerical code until getting very 
stable results, which do not depend on $\a$. In practice, we have used $\a=10^{-4}$ in all numerical calculations}:
\begin{align}
&f=R+\beta |R|^N \quad &&\to \quad f=R+\beta |R|^N+\alpha R^2: \\ \nonumber
 &f''=\beta N(N-1)|R|^{N-2}
 \quad &&\to \quad f''=\beta N(N-1)|R|^{N-2}+2\alpha \,.
\end{align}

In the previous section we have explored the regime where 
the higher-curvature term presents a small correction 
to the Einstein term,
$\beta R^N \lesssim R$,
i.e. the weak coupling limit.
Now let us describe what happens beyond the weak coupling limit.
It is well-known that in the strong coupling limit there is 
another type of dynamics described by the stable solution with Big Rip singularity (\cite{COA}-\cite{fRGB}),
\[
a\propto (t_{BR}-t)^{\frac{2N^2-3N+1}{2-N}}, \quad H\sim \frac{1}{t_{BR}-t}\,,
\]
where $t_{BR}$ denotes the time of singularity.

As expected from the analysis in the Einstein frame (\ref{EF}) for $N>2$
we have two asymptotics: an oscillatory solution and a run-away solution.
This expectation is confirmed by our numerical procedure, see Fig.\ref{Fig:0}.
We have obtained that 
low-curvature initial conditions $R_0(H_0,\dot{H}_0)\to 0$ also lead to an oscillatory solution even in the case of strong couplings $\beta \gtrsim 1$.
The oscillatory solution  found numerically for strong couplings
exhibits the same features with that found analytically in the 
weak coupling limit, allowing us to identify them.
Admirably, the behaviour of the energy densities of matter and HCGC
depend on the interplay 
between the equation-of-state 
parameter for ordinary matter $\gamma$ and the parameter $\gamma_{crit}$,
(defined by \eqref{gmcrit})
in same manner as it has been discussed in Subsec.(\ref{Res}).
If $\gm>\gm_{crit}$ HCGC dominate over the matter contribution at late times 
and \textit{vice versa} for $\gm<\gm_{crit}$, see Fig.\ref{Fig:5}.
In the first case the matter energy density decays with time faster than in the
last case in agreement with \eqref{dom_osc}.
The time-behaviour of densities ratio is plotted in  Fig.\ref{Fig:3} and illustrates main points of the analytical study presented in Subsec.\ref{Res}.
For $\gm<\gm_{crit}$ the Hubble rate asymptotically tends to the 
GR value given by $ H^2=\r$ while
for $\gm<\gm_{crit}$ we have oscillations of $H^2/\r$ 
with an increasing amplitude.

\begin{figure}[!h]
\begin{center}
\includegraphics[width=0.45\textwidth]{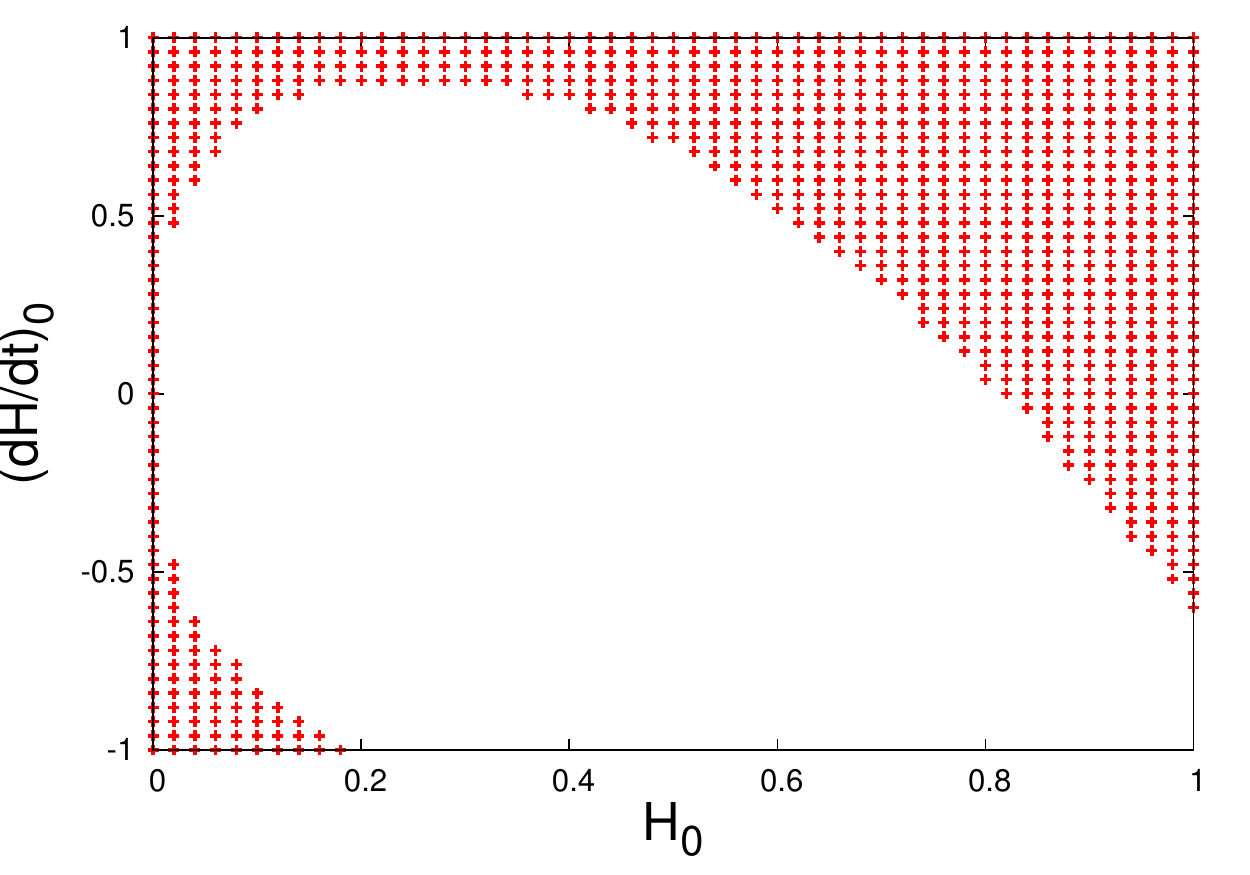}
\includegraphics[width=0.45\textwidth]{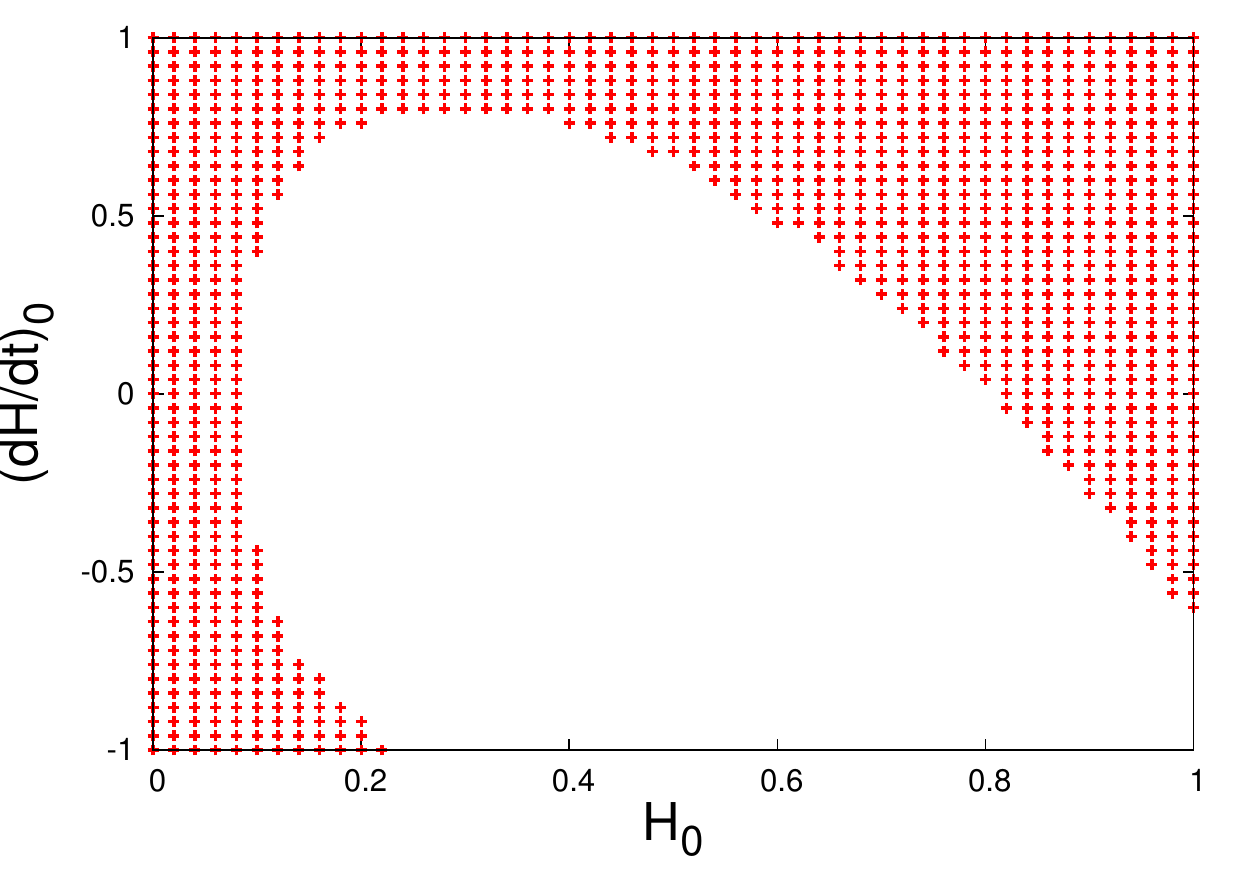}
\includegraphics[width=0.45\textwidth]{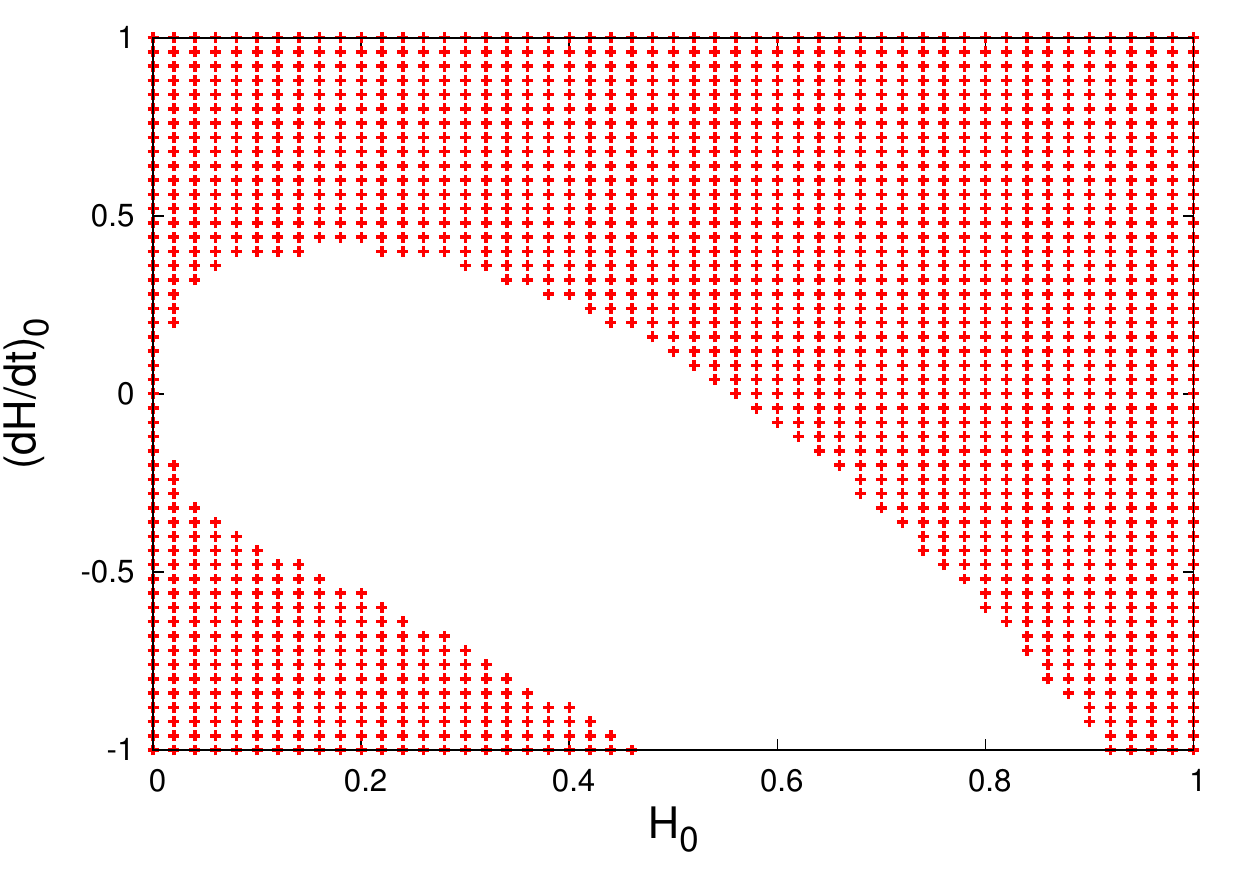}
\includegraphics[width=0.45\textwidth]{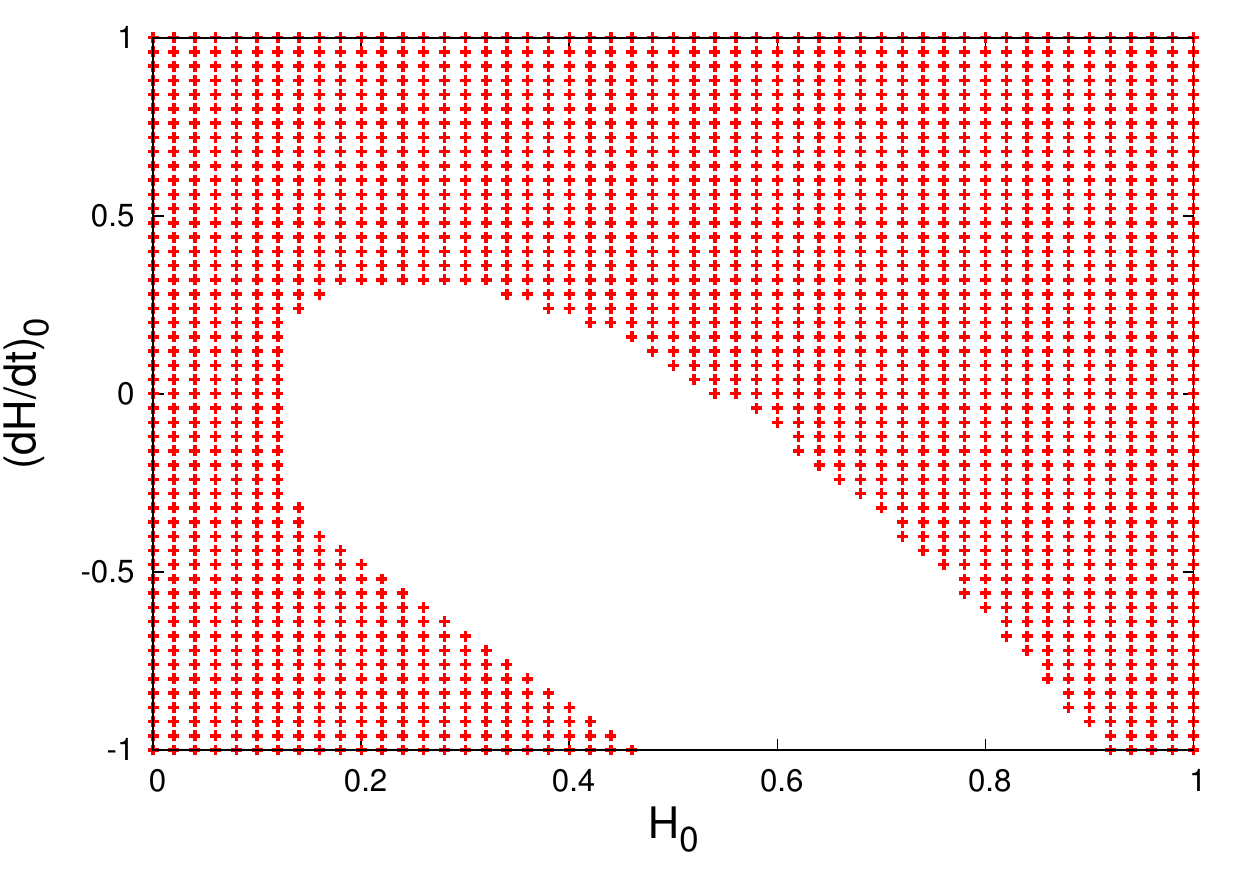}
\includegraphics[width=0.45\textwidth]{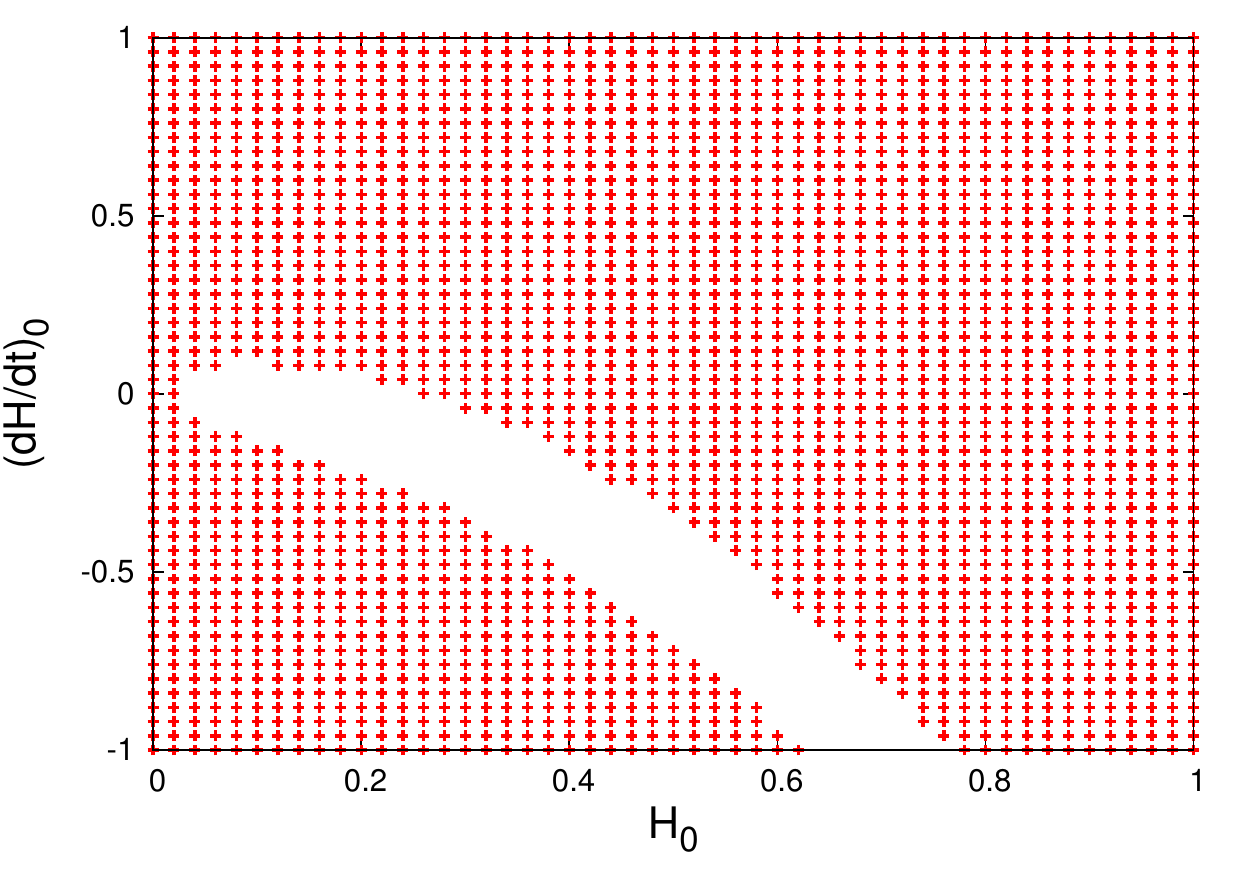}
\includegraphics[width=0.45\textwidth]{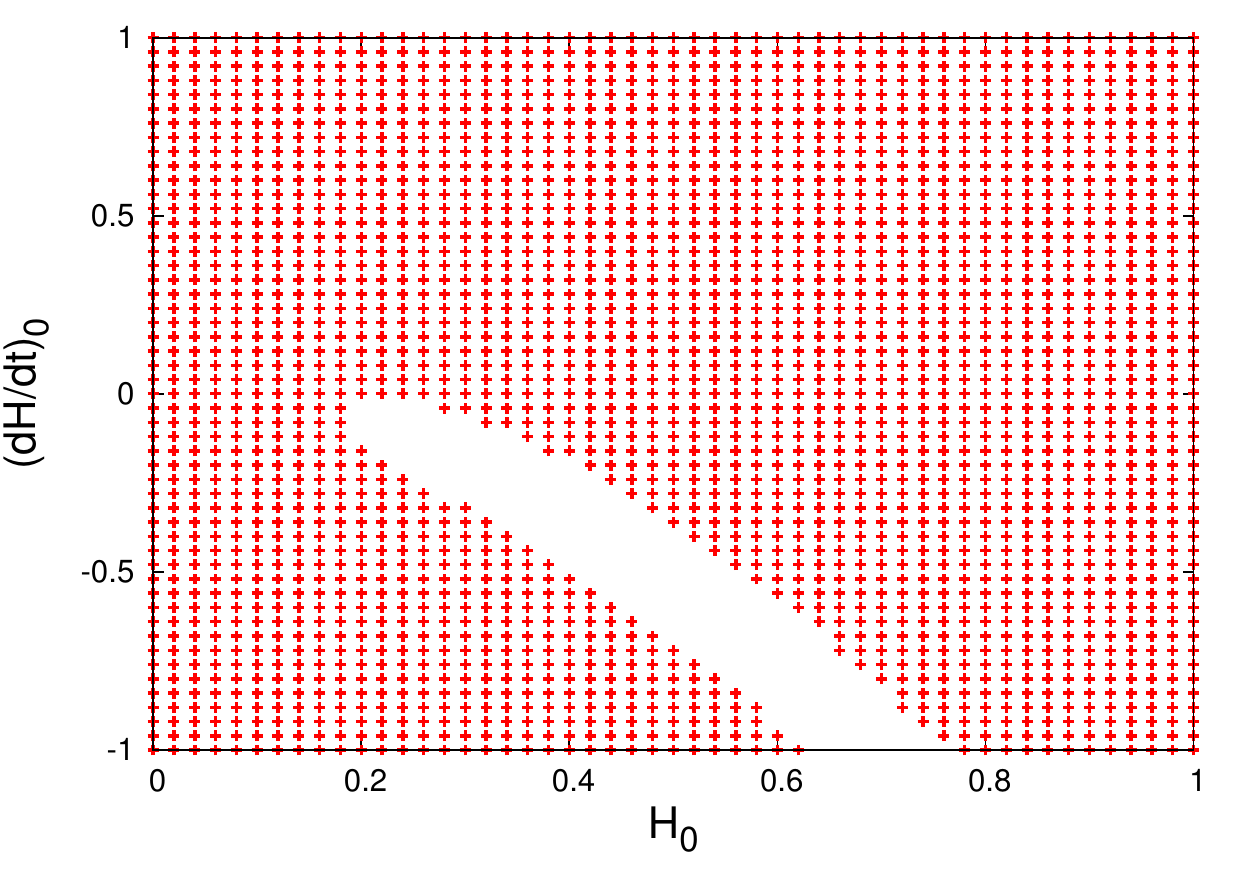}
\caption{Basin of attraction for the Big Rip solution. 
$H_0$ and $(dH/dt)_0$ are initial conditions for eq. (\ref{00f2}).
Red points mark initial data, leading to the Big Rip singularity.
Continuous white area corresponds to initial conditions, leading to 
an oscillating asymptotic.
 The model parameters are: $\alpha=10^{-4}, N=4$, $\beta=10^{-3}$(upper panels), $\beta=10^{-2}$ (middle panels), $\beta=1$ (lower panels). 
On the left panels the initial density value for dust ($\gm=1$) is $\rho_{0}=0.03$; on the right panels $\rho_{0}=0.3$.
\label{Fig:1}
}
\end{center}
\end{figure}

Now let us define which initial conditions lead to each asymptotic.
The basins of attraction 
in the initial condition space $(H_0,\dot{H}_0)$ for two slices of initial matter 
density $\r_0$
are presented in  Fig.\ref{Fig:1} for 
a set of couplings $\beta$\footnote{As a working example we choose the matter to be ordinary dust 
with $\gm=1$. We checked that the attraction basins 
for other types of 
fluids with $\gm$ different from $1$
are qualitatively the same to those presented here.}
For numerics we have chosen one particular case of $N=4$.
We clearly see that increasing the coupling $\beta$ 
the basin of attraction for an oscillatory solution
becomes  narrower (though non-vanishing) and concentrates near the parabola $R_0(H_0,\dot{H}_0)=12H^2_0+6\dot{H}_0=0$, while the basin of attraction
for the Big-Rip solution covers more and more space.

\section{Conclusions}
\label{conclusions}

In this paper we have considered an oscillatory regime in $f(R)$ FRW cosmology with power-law functions $f=R+\beta R^N$. The oscillations in general 
appeared to be anharmonic (except the case of $N=2$)
and have an effective equation of state $p_{f(R)}=(\gamma_{crit}-1)\rho_{f(R)}$ with $\gamma_{crit}=2N/(3N-2)$, meaning that for any $N>2$ these oscillations
have negative effective pressure. On the other hand, $\gamma_{crit} \to 2/3$ for $N \to \infty$, being always bigger that $2/3$, so these oscillations can not cause an accelerated expansion.

The case of $N=2$ is an exceptional one in the family of power-law $f(R)$ for two further
reasons: 
First,the coefficient at the highest derivative term in the equation of motion never vanishes in an
expanding Universe. On the contrary, it may vanish for $N>2$ giving an example of a 
"non-standard singularity" with finite curvature and diverging curvature time derivative. We show that such
a "singularity" 
(in a more general sense that the usual one, because the curvature does not 
diverge) is traversable and can be removed by small change in the form of $f(R)$ at least for an even $N$ (or for an arbitrary $N>1$, if we consider the theory with $f=R+\beta |R|^N$).  

And second, there is a stable phantom-like asymptotic in the $N>2$ case which is absent
for $N=2$. Initial data leading to this Big Rip regime is located in a zone of large initial curvature. If a trajectory starts from low curvature initial conditions (what value of
a curvature is low enough depends on the coupling constant $\beta$ at the $R^N$ term) it falls into
the oscillation regime.

\section*{Acknowledgements}
Authors are grateful to Alexandr Panin, Sergey Sibiryakov and Anna Tokareva for useful discussions. M.I. dedicates his work to Evgenia Dueva, who always insists on perfectness.
This work was supported in part by the Grants of the President of Russian Federation MK-1754.2013.2 and NS-5590.2012.2 (MI), the RFBR grants 12-02-31708 (M.I.), 13-01- 00200 (E.B.), 12-01-00387 (E.B.), 11-02-00643 (A.T.), 14-02-00894 (A.T. and M.I.) the grant of the Ministry of Education and Science No 8412. (FAE program by government contract 16.740.11.0583) (M.I.) and by the Dynasty Foundation (M.I.).

\appendix

\section{Generalised trigonometric functions}
\label{App1}

\begin{figure}[h]
\centering
  \hfill \includegraphics[width=0.4\textwidth]{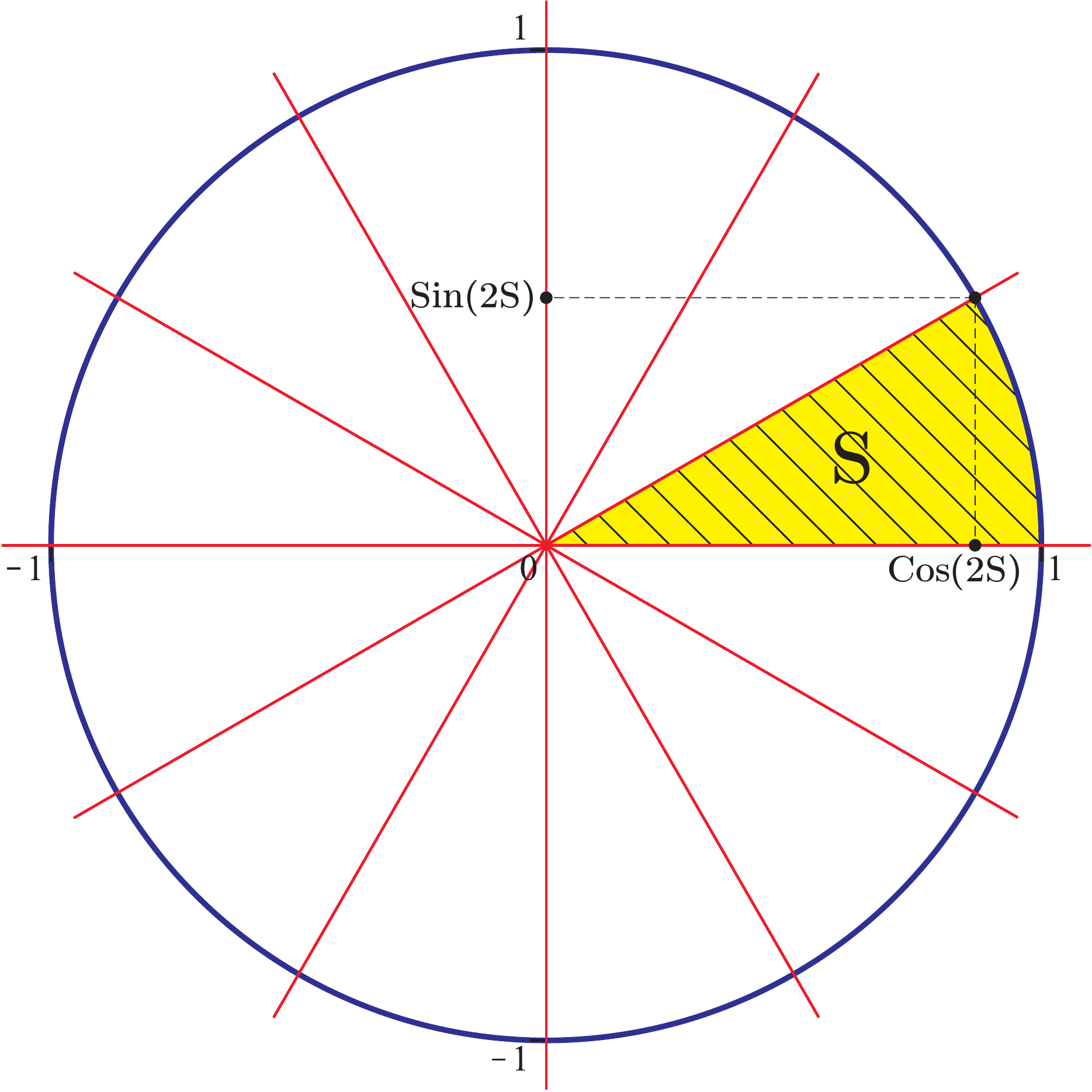}
  \hfill
  \includegraphics[width=0.4\textwidth]{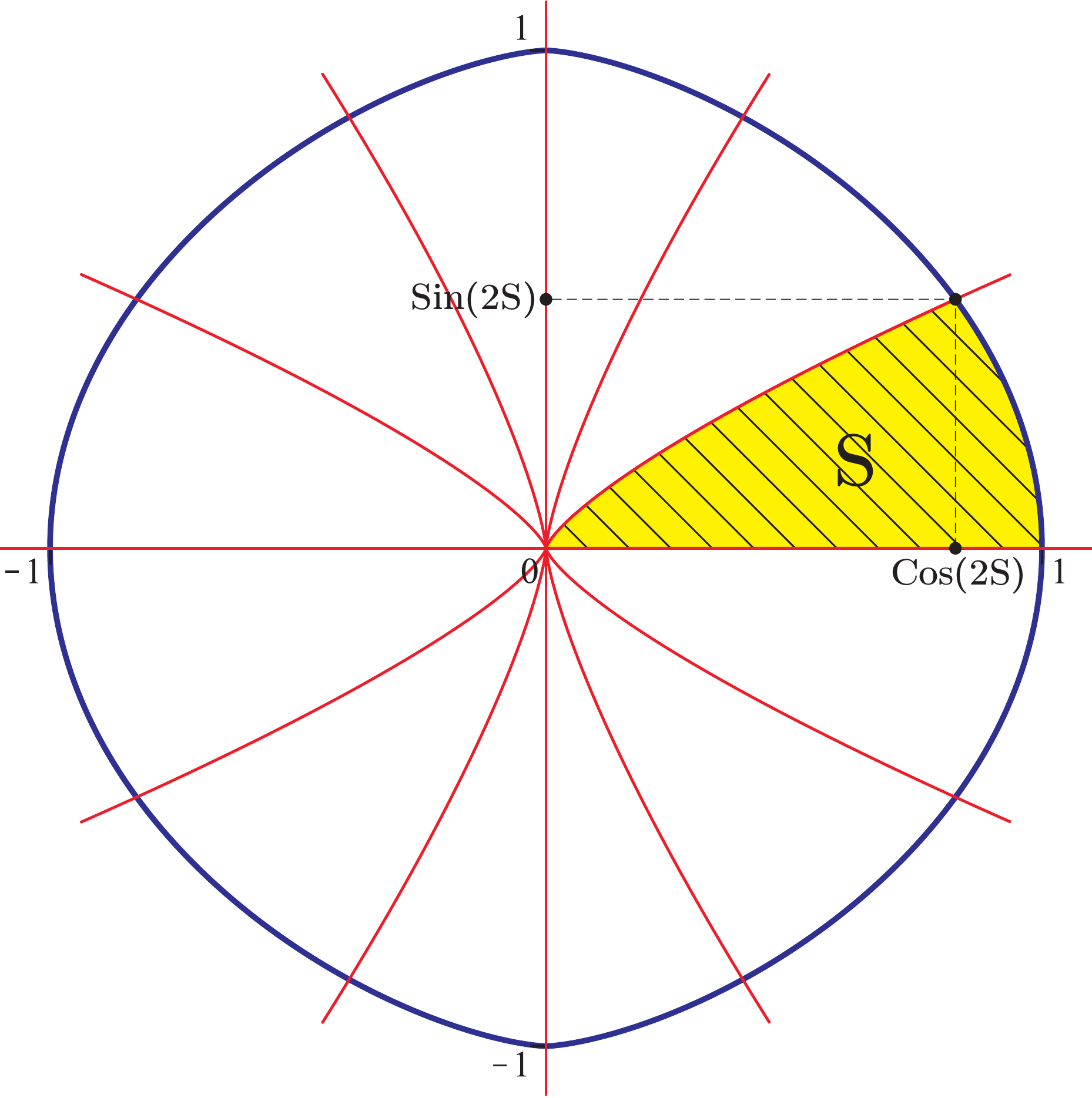} \hfill\ms \\
  \hfill \parbox{.4\textwidth}{\caption{Standard trigonometric circle with the unit radius.} \label{fig Cos-Sin_0}}
  \hfill \parbox{.4\textwidth}{\caption{The curve $\Gamma$ (see \eqref{formula Gm}) which has been used to define generalized trigonometric functions.} \label{fig Cos-Sin_1}}
  \hfill\ms \\[-3ex]
\end{figure}

Everybody knows that ordinary trigonometric functions such as $\sin x$ and~$\cos x$
may be defined by a trigonometric circle with the unit radius. This method 
is pictured on Fig.~\ref{fig Cos-Sin_0}, where $S$ denotes an area of the shaded region.
We will use similar method to define generalised sine and cosine.

We remind the reader that the area~$S$ relates to the polar angle~$\vph$ simply as $\vph = 2\, S$. Thus, 
\begin{equation}\label{cossin0}
  \cos \vph = \cos [2\, S(\vph)], \qquad \sin \vph = \sin [2\, S(\vph)].
\end{equation}

Relation~\eqref{cossin0} may be expanded to the case of arbitrary~$\vph \in \RR$: if~$\vph = \vph_0 + 2\, \pi\, k$, where $\vph_0 \in [0, 2\, \pi)$, $k \in \mathbb{Z}$, then~$S = S_0 + S_{\mathrm {cir}}\, k = S_0 + \pi\, k$, where~$S_0$ is the area of
the circle, corresponding to~$\vph = \vph_0$, and $S_{\mathrm{cir}}$ is the total
area of the circle.

Now consider a plane curve~$\Gm$, defined by the following equation:
\begin{equation}\label{def Gm}
  \Gm: {|x|}^a + {|y|}^b = 1,
\end{equation}
where $a, b > 1$ (The case of $a = 3/2$, $b = 2$ is presented in Fig.~\ref{fig Cos-Sin_1}).
To parametrize this curve, let us go to the 
following generalised polar coordinates $\rho$ and~$\vph$:
\begin{align*}
  x &= \rho^{2/a}\, {|{\cos{}} \vph|}^{2/a}\, \sign \cos \vph,
\\
  y &= \rho^{2/\Ch[\scs]{b}{a}}\, {|\Ch{\sin}{\cos{}} \vph|}^{2/\Ch[\scs]{b}{a}}\, \sign \Ch{\sin}{\cos{}} \vph.
\end{align*}

In such coordinates the equation (\ref{def Gm}) reads as 
$\rho \equiv 1$, and going back
to~$x$ and~$y$ one has:
\begin{equation}\label{formula Gm}
  \Gm: \Bigg\{
  \begin{aligned}
    x &= {|{\cos{}} \vph|}^{2/a}\, \sign \cos \vph,
  \\
    y &= {|\Ch{\sin}{\cos{}} \vph|}^{2/\Ch[\scs]{b}{a}}\, \sign \Ch{\sin}{\cos{}} \vph.
  \end{aligned}
\end{equation}

Further, we will consider the parameter $\vph$ as a function of 
a doubled area~$2S$ of the region~$\Dl\Om$, enclosed by the curve~$\vph = \const$, the $x$-axis and the curve~$\rho = 1$
(see~Fig.~\ref{fig Cos-Sin_1}):
\begin{equation}\label{vph of t}
  \vph = \vph(t),\ \text{where}\ t = 2\, S.
\end{equation}

Note that in the Cartesian coordinates the curve~$\vph = \const$ may be presented 
in the following way (depending on the range of~$\vph$):
\begin{enumerate}
  \item
  if~$\cos \vph \ne 0$,
  \[
    y = {|x|}^{a/b}\, |\tg \vph|^{2/b}\, \sign \sin \vph, \quad x \in [0, \mp\infty) \quad \text{for}\, \cos \vph \lessgtr 0;
  \]
  \item
  if~$\cos \vph = 0$, 
  \begin{equation}\label{cos vph = 0}
    x = 0, \quad y \in [0, \mp\infty) \quad \text{for}\, \sin \vph \lessgtr 0.
  \end{equation}
\end{enumerate}

Several curves of the 	
assemblage~$\vph = \const$ are presented in Fig.~\ref{fig Cos-Sin_1}. 
Clearly, the~$\vph$ and~$S$ 
are in one-to-one correspondence (so~$S$ and~$t$ may be considered as a function of~$\vph$) and this correspondence may be expanded to arbitrary real~$\vph$ by the agreement that value
\begin{equation}\label{vph & S}
  \vph = \vph_0 + 2\, \pi\, k,
\end{equation}
where $\vph_0 \in [0, 2\, \pi)$, $k \in \mathbb{Z}$, corresponds to the value 
\begin{equation}\label{vph & S'}
  S = S_0 + S_{\mathrm {cur}}\, k,
  \tag{\ref{vph & S}$'$}
\end{equation}
where~$S_{\mathrm{cur}}$~is the area~$\Om$, enclosed by the curve~$\Gm: {|x|}^a + {|y|}^b = 1$; $S_0$~is the area~$\Dl\Om$~of the part of~$\Om$, corresponding to the value~$\vph = \vph_0$.

Next, we find a relation between~$\vph$ and~$t$ by calculating the 
transformation
Jacobian:
\begin{equation}\label{area}
  S = \iint\limits_{\Dl\Om} dx\, dy = \iint\limits_{\Dl\om} \Bigg| \frac{D(x,y)}{D(\rho,\al)} \Bigg|\, d\rho\, d\al, \quad \text{where} \quad \Dl\om = \Big\{ (\rho,\al) \mid \rho \in [0,1], \al \in [0,\vph] \Big\},
\end{equation}
\begin{equation}\label{Jacobian}
  \frac{D(x,y)}{D(\rho,\vph)} = \tfrac4{a\,b}\, \rho^{\frac2a + \frac2b -1}\, {|{\cos{}} \vph|}^{\frac2a-1}\, {|{\sin{}} \vph|}^{\frac2b-1}.
\end{equation}
Plugging~\eqref{Jacobian} into the eq.~\eqref{area} for~$S$ one finds:
\begin{multline}\label{formula S}
  S = \tfrac4{a\,b} \intl01 \rho^{\frac2a + \frac2b -1}\, d\rho \intl0\vph {|{\cos{}} \al|}^{\frac2a-1}\, {|{\sin{}} \al|}^{\frac2b-1}\, d\al =
\\
  = \tfrac4{a\,b}\, \tfrac{a\,b}{2\,(a+b)}\, \rho^{\frac{2\,(a+b)}{a\,b}} \Big|_0^1\, \intl0\vph {|{\cos{}} \al|}^{\frac2a-1}\, {|{\sin{}} \al|}^{\frac2b-1}\, d\al = \tfrac2{a+b} \intl0\vph {|{\cos{}} \al|}^{\frac2a-1}\, {|{\sin{}} \al|}^{\frac2b-1}\, d\al.
\end{multline}

Using~$t = 2\, S$, we obtain the final relation between~$t$ and~$\vph$:
\begin{equation}\label{t formula}
  t = \tfrac4{a+b} \intl0\vph {|{\cos{}} \al|}^{\frac2a-1}\, {|{\sin{}} \al|}^{\frac2b-1}\, d\al.
\end{equation}

The relation~\eqref{t formula} (for fixed~$a$ and~$b$) defines 
one-to-one correspondence between $t$ and~$\vph$. 
Let us denote its solution with
respect to~$\vph$ as~$\vph(a,b;t)$. 
Clearly, $\vph(a,b;t)$ may not be expressed in elementary functions for
arbitrary values of~$a$ and~$b$.
Using the function~$\vph(a,b;t)$, let us introduce sine and cosine of order~${\bf p} = \{ a,b \}$.
\begin{Def}\label{Def Cos Sin}
Functions 
  \begin{equation}\label{def Cos Sin}
    \begin{aligned}
      \Cos(a,b;t) &= {|{\cos{}} \vph(a,b;t)|}^{2/a}\, \sign \cos \vph(a,b;t),
    \\
      \Ch{\Sin}{\Cos}(a,b;t) &= {|\Ch{\sin}{\cos{}} \vph(a,b;t)|}^{2/\Ch[\scs]{b}{a}}\, \sign \Ch{\sin}{\cos{}} \vph(a,b;t)
    \end{aligned}
  \end{equation}
  are called the cosine and the sine of order \textbf{p} $=\{a,b\}$ respectively.
\end{Def}

We will omit $a$ and $b$ in the places, where we use a sine and a cosine
of the same order, i.e. to write $\Cos t$, $\Sin t$ and~$\vph(t)$ instead of~$\Cos(a,b;t)$, $\Sin(a,b;t)$ and~$\vph(a,b;t)$. 
Moreover, we will call the functions~$\Cos t$ 
and~$\Sin t$ just as generalized cosine and sine.
As it may be seen from~\eqref{def Cos Sin} and~\eqref{formula Gm}, functions 
$\Cos t$ and~$\Sin t$ represent correspondingly coordinates $x$ and $y$ of a crossing point of the curves $\Gm$ and~$\vph = \const$, enclosing the area $S = t/2$ (see Fig.~\ref{fig Cos-Sin_1}).

Let us find differential equations for $\Cos(a,b;t)$ and~$\Sin(a,b;t)$.
From the definition~\eqref{def Cos Sin} we have (a prime denotes derivative with respect to $t$)
\begin{equation}\label{Cos'Sin'1}
  \begin{aligned}
    {(\Cos t)}' &= {}- \tfrac2a\, {|{\cos{}} \vph(t)|}^{\frac2a-1}\, [\Ch{\sin}{\cos{}} \vph(t)]\, \vph'(t),
  \\
    {(\Ch{\Sin}{\Cos{}} t)}' &= \Ch{{}+{}}{{}-{}} \tfrac2{\Ch[\scs]{b}{a}}\, {|\Ch{\sin}{\cos{}} \vph(t)|}^{\frac2{\Ch[\sss]{b}{a}}-1}\, [\cos \vph(t)]\, \vph'(t).
  \end{aligned}
\end{equation}

The derivative~$\vph'(t)$ may be expressed from~\eqref{t formula}:
\begin{equation}\label{vph'}
  \vph'(t) = \tfrac{a+b}4\, {|{\cos{}} \vph(t)|}^{1-\frac2a}\, {|{\sin{}} \vph(t)|}^{1-\frac2b}.
\end{equation}

Plugging~\eqref{vph'} into~\eqref{Cos'Sin'1}, we have
\begin{align}\label{Cos'}
  {(\Cos t)}' &= {}- \tfrac{a+b}{2\,a}\, {|\Ch{\sin}{\cos{}} \vph(t)|}^{\frac2{\Ch[\sss]{b}{a}} (\Ch[\scs]{b}{a} - 1)}\, \sign \Ch{\sin}{\cos{}} \vph(t) = {}- \tfrac{a+b}{2\,a}\, {|\Ch{\Sin}{\Cos{}} t|}^{\Ch[\scs]{b}{a} - 1}\, \sign \Ch{\Sin}{\Cos{}} t,
\\[1ex]\label{Sin'}
  {(\Ch{\Sin}{\Cos{}} t)}' &= \Ch{{}+{}}{{}-{}} \tfrac{a+b}{2\,b}\, {|{\cos{}} \vph(t)|}^{\frac2a (a-1)}\, \sign \cos \vph(t) = \Ch{{}+{}}{{}-{}} \tfrac{a+b}{2\,b}\, {|{\Cos{}} t|}^{a-1}\, \sign \Cos t.
\end{align}

Using the identity (see.~\eqref{def Cos Sin},~\eqref{def Gm}):
\begin{equation}\label{Trig equiv}
  {|{\Cos{}} t|}^a + {|{\Sin{}} t|}^b \equiv 1,
\end{equation}
 we obtain
\begin{align}\label{sin to cos}
  {|\Ch{\Sin}{\Cos{}} t|}^{\Ch[\scs]{b}{a} - 1} &\equiv {\big( 1 - {|{\Cos{}} t|}^a \big)}^{\frac{\Ch[\sss]{b}{a} - 1}b},
\\ \label{cos to sin}
  {|{\Cos{}} t|}^{a-1} &\equiv {\big( 1 - {|\Ch{\Sin}{\Cos{}} t|}^{\Ch[\scs]{b}{a}} \big)}^{\frac{a-1}a}.
\end{align}

Clearly, from~\eqref{Cos'} and~\eqref{sin to cos} one has
\[
  {\big( \tfrac{2\,a}{a+b} \big)}^{\frac b{b-1}}\, {|{\Cos' t}|}^{\frac b{b-1}} + {|{\Cos t}|}^a = 1.
\]
Using the fact that~$\Cos 0 = 1$ (see~\eqref{def Cos Sin} and~\eqref{t formula}), we conclude that the function~$z = \Cos t$ is a solution of the following Cauchi problem:
\begin{subequations}
\begin{gather}\label{Cos equat 1}
  {\big( \tfrac{2\,a}{a+b} \big)}^{\frac b{b-1}}\, {|\dot z|}^{\frac b{b-1}} + {|z|}^a = 1,
\\ \label{Cos init 1}
  z(0) = 1.
\end{gather}
\end{subequations}

Perming similar actions, using \eqref{Sin'},~\eqref{cos to sin} and the fact
that~$\Sin 0 = 0$ (see~\eqref{def Cos Sin},~\eqref{t formula}),
we obtain the following initial problem for the function~$z = \Sin t$:
\begin{subequations}
\begin{gather}\label{Sin equat 1}
  {\big( \tfrac{2\,b}{a+b} \big)}^{\frac a{a-1}}\, {|\dot z|}^{\frac a{a-1}} + {|z|}^b = 1,
\\ \label{Sin init 1}
  z(0) = 0.
\end{gather}
\end{subequations}

Equation~\eqref{Cos equat 1} for~$\Cos t$ is not resolved with respect to a higher derivative. To get such kind of equation, let us take one more derivative of~\eqref{Cos equat 1}:
\begin{equation}\label{Cos equat}
  {\big( \tfrac{2\,a}{a+b} \big)}^{\frac b{b-1}}\, \tfrac b{b-1}\, {|\dot z|}^{\frac1{b-1}}\, (\sign \dot z)\, \ddot z + a\, {|z|}^{a-1}\, (\sign z)\, \dot z = 0.
\end{equation}

Using eq.~\eqref{Cos init 1},~$\Cos' 0 = 0$ (see.~\eqref{Cos'Sin'1} and~\eqref{t formula}), 
we get the following second-order Cauchi problem for $z = \Cos t$:
\begin{gather*}
  \tfrac b{b-1}\, {\big( \tfrac{2\,a}{a+b} \big)}^{\frac b{b-1}}\, {|\dot z|}^{\frac{2-b}{b-1}}\, \ddot z + a\, {|z|}^{a-2}\, z = 0,
\\
  z(0) = 1, \quad \dot z(0) = 0.
\end{gather*}

Similarly, from \eqref{Sin equat 1},~\eqref{Sin init 1},~$\Sin' 0 = (a+b)/(2\, b)$ (see.~\eqref{Cos'Sin'1}, \eqref{t formula} and ~\eqref{vph'}), we find the initial problem for~$z = \Sin t$:
\begin{gather}
\label{z''}
  \tfrac a{a-1}\, {\big( \tfrac{2\,b}{a+b} \big)}^{\frac a{a-1}}\, {|\dot z|}^{\frac{2-a}{a-1}}\, \ddot z + b\, {|z|}^{b-2}\, z = 0,
\\
  z(0) = 0, \quad \dot z(0) = \tfrac{a+b}{2\,b}.
\end{gather}

Next, let us study some useful features of~$\Cos t$ and~$\Sin t$.
First of all, from the initial geometrical 
interpretation (see Fig.~\ref{fig Cos-Sin_1})
one concludes that $\Cos t$ and~$\Sin t$ are periodic functions with the period~$T = 2\, S_{\mathrm {cur}}$, where~$S_{\mathrm {cur}}$~is the square of area~$\Om$, enclosed by the curve~$\Gm$:
\begin{equation}\label{Periodicity Cos Sin}
  \Cos(a,b;t+T) \equiv \Cos(a,b;t), \qquad \Sin(a,b;t+T) \equiv \Sin(a,b;t).
\end{equation}

To find~$T$ one should use the fact that the expression~\eqref{formula S}
for area $\Dl\Om$ in between the curve $\vph = \const$ and $X$-axis coincides
with the $\Om$ for~$\vph = 2\, \pi$:
\begin{equation}\label{formila_1 T}
  T = 2\, S_{\mathrm {cur}} = \tfrac4{a+b}\, \intl0{2\,\pi} {|{\cos{}} \al|}^{\frac2a-1}\, {|{\sin{}} \al|}^{\frac2b-1}\, d\al.
\end{equation}

The period $T$ may be expressed 
in terms of the Euler's Gamma $\Gm(x)$ and Beta $B(x,y)$ functions. From~\eqref{formila_1 T} one writes
\begin{multline*}
  \intl0{2\,\pi} {|{\cos{}} \al|}^{\frac2a-1}\, {|{\sin{}} \al|}^{\frac2b-1}\, d\al = 4\, \intl0{\pi/2} {\big( {\cos^2{}} \al \big)}^{\frac1a-1}\, {\big( {\sin^2{}} \al \big)}^{\frac1b-1}\, \cos\al\, \sin\al\, d\al =
\\
  = \intl{\pi/2}0 {\big( \tfrac{1 + \cos 2\al}2 \big)}^{\frac1a-1}\, {\big( \tfrac{1 - \cos 2\al}2 \big)}^{\frac1b-1}\, d\cos 2\al = \intl{-1}{+1} {\big( \tfrac{1 + u}2 \big)}^{\frac1a-1}\, {\big( \tfrac{1 - u}2 \big)}^{\frac1b-1}\, du.
\end{multline*}

Using~\eqref{formila_1 T} and performing the change of variable $u = 2\,v - 1$, we get the final result:
\begin{equation}
\label{Tvalue}
  T = \tfrac8{a+b}\, \intl01\, v^{\frac1a-1}\, {(1-v)}^{\frac1b-1}\, dv = \tfrac8{a+b}\, B(\tfrac1a, \tfrac1b) = \tfrac8{a+b}\, \tfrac{\Gm(\frac1a)\, \Gm(\frac1b)}{\Gm(\frac1a + \frac1b)}.
\end{equation}
For $a = b = 2$ the period $T$ reduces to well-known value
\[
  T = 2\, \tfrac{\Gm(1/2)\, \Gm(1/2)}{\Gm(1)} = 2\, \tfrac{\sqrt{\pi}\, \sqrt{\pi}}{1} = 2\, \pi,
\]
as expected from the fact that $\Cos t$ and ~$\Sin t$ of order ${\bf p} = \{ 2,2 \}$ are ordinary trigonometric functions (see~\eqref{def Cos Sin} and~\eqref{t formula}):
\[
  \Cos(2,2;t) = \cos t, \qquad \Sin(2,2;t) = \sin t.
\]
Finally, from the Def.~\ref{Def Cos Sin} and $\vph(-t) = -\vph(t)$ (as a consequence of
\eqref{vph & S}, \eqref{vph & S'} and our agreement to expand $\vph$ and~$S = 2\,t$ to all real numbers) we obtain evenness of~$\Cos t$ and oddness of~$\Sin t$:
\[
  \Cos(-t) = \Cos t, \qquad \Sin(-t) = {}- \Sin t.
\]

Now we are equipped to solve the eq.\eqref{1eqH}.
Performing the change of variable
\be 
\tilde{\psi}=\tfrac{4(N-1)}{3N-2}(C_1^2-C_2)^{\frac{2-N}{2N}}\tau+\text{const}\,, 
\ee
our equation transforms to:
\be 
\nu^N|z^{\prime}|^{N}+z^2=1\,,\quad \text{where}\quad z\equiv \frac{H-C_1}{\sqrt{C_1^2-C_2}}\,,\quad z^{\prime}=\frac{dz}{d\tilde{\psi}}\,,\quad \frac{4(N-1)}{3N-2}\equiv \nu\,.
\ee
Comparing this with (\ref{Sin equat 1}) one finds solutions of the eqs.(\ref{1eqH}),(\ref{eq2}):
\begin{align}
\label{eqtp}
H&=C_1+\sqrt{C_1^2-C_2}\;\text{Sin}(\tfrac{N}{N-1},2;\tilde{\psi})\,,\quad \text{where}\quad \tilde{\psi}=\nu(C_1^2-C_2)^{\tfrac{2-N}{2N}}\tau+\text{const}\,,\\
Q=H_{\tau}&=\nu\sqrt[N]{C_1^2-C_2}\;\text{Sin}^{\prime}(\tfrac{N}{N-1},2;\tilde{\psi})\,=\nonumber \\
&=\sqrt[N]{C_1^2-C_2}\;\sqrt[N]{1-\Sin^2(\tfrac{N}{N-1},2;\tilde{\psi})}\,\sign\Sin^{\prime}(\tfrac{N}{N-1},2;\tilde{\psi})\,.
\end{align}
Introducing new notations and absorbing constant $\nu$ in the independent variable $\tp$ we get:
\begin{align}
\label{defSinN}
\text{Sin}({\displaystyle{\tiny \tfrac{N}{N-1}}},2;\tilde{\psi})&\equiv\sin_N(\psi)\,,
\quad \text{where} \quad \psi\equiv (C_1^2-C_2)^{\frac{2-N}{2N}}\tau+\text{const}\,,\\
\text{Cos}({\displaystyle{\tiny \tfrac{N}{N-1}}},2;\tilde{\psi})&\equiv\cos_N(\psi)\,,\\
\sign\Sin^{\prime}(\tfrac{N}{N-1},2;\tilde{\psi})&=\sign Q(\tau)\equiv \Sg_N(\psi)\,,\\
\label{sinNprime}
\frac{\pd}{\pd\psi}\sin_N(\psi)&=\Sg_N(\psi)\sqrt[N]{1-\sin_N^2(\psi)}\,;
\end{align}
and write down final solution of an unperturbed system:
\begin{subequations}
\begin{align}
H&=H(C_1,C_2,\psi)\equiv C_1+\sqrt{C_1^2-C_2}\sin_{N}(\psi)\,,\\
Q&=Q(C_1,C_2,\psi)\equiv \Sg_N(\psi)\sqrt[N]{C_1^2-C_2}\sqrt[N]{1-\sin^2_N(\psi)}\,,\\
\r&=\r(C_1,C_2,\psi)\equiv C_2\,.
\end{align}
\end{subequations}
Finally, since $\tp=\nu\psi$ (see \eqref{eqtp},\eqref{defSinN}), the period $T_0$ of the function $\sin_N(\psi)$
is $\nu$ times smaller than the period $T$ of the function $\Sin(\tfrac{N}{N-1},2;\tilde{\psi})$:
\be
\label{TsinN}
T_0=\tfrac{1}{\nu}T=2B(\tfrac{1}{2},\tfrac{N-1}{N})\,. 
\ee

\end{document}